\documentclass[prd,10pt,aps,onecolumn,nofootinbib]{revtex4-2}

\usepackage{latexsym}
\usepackage{hyperref}
\usepackage{amssymb}
\usepackage{epsfig,amsmath,graphics}
\usepackage{color}
\usepackage{dcolumn}
\usepackage[normalem]{ulem}
\usepackage{xcolor}
\usepackage{placeins}
\usepackage{braket}
\usepackage[capitalize]{cleveref}
\usepackage{multirow}
\usepackage{float}

\usepackage{csquotes}

\definecolor{linkcolor}{rgb}{0.0, 0.28, 0.67}
\hypersetup{colorlinks=true,
	linkcolor=black,
	citecolor=linkcolor,
	urlcolor=linkcolor,
	linktocpage=true
}

\usepackage{xspace}
\newcommand{\wzdrnu}{WZDR$\nu$\xspace}

\newcommand{\Neff}{\ensuremath{N_{\rm eff}}}

\newcommand{\Nfld}{N_{\rm IDR}}

\newcommand{\LCDM}{\ensuremath{\Lambda{\rm CDM}}\xspace}
\newcommand{\nuSM}{\nu}
\newcommand{\DS}{\ensuremath{\mathrm{DS}}}
\newcommand{\rhoDS}{\rho_\DS}

\newcommand{\TDS}{T_\DS}

\newcommand{\nudark}{\ensuremath{{\nu_{d}}}}
\newcommand{\mdark}{m_{\nu d}}

\newcommand{\alphad}{\alpha_{d}}
\newcommand{\Tequil}{T_{\rm equil}}

\newcommand{\Gammaph}{\Gamma_{\text{ph.}}}

\newcommand{\Niur}{N_\mathrm{FS}}

\newcommand{\ffs}{free-streaming+self-interacting\xspace}

\begin{document}
\title{Cosmological probes of Dark Radiation from Neutrino Mixing}
	\author{Itamar J. Allali}
 \email{itamar\_allali@brown.edu}
        \affiliation{Department of Physics, Brown University, Providence, RI 02912, USA}
	\affiliation{Institute of Cosmology, Department of Physics and Astronomy, Tufts University, Medford, MA 02155, USA}
	\author{Daniel Aloni}
 \email{alonidan@bu.edu}
	\affiliation{Physics Department, Boston University, Boston, MA 02215, USA}
	\affiliation{Department of Physics, Harvard University, Cambridge, MA 02138, USA}
	\author{Nils Sch\"oneberg}
 \email{nils.science@gmail.com}
	\affiliation{Institut de Ci\`encies del Cosmos, Universitat de Barcelona, Mart\'{\i} i Franqu\`es 1, Barcelona 08028, Spain}

	\begin{abstract}
Models of stepped dark radiation have recently been found to have an important impact on the anisotropies of the cosmic microwave background, aiding in easing the Hubble tension. In this work, we study models with a sector of dark radiation with a step in its abundance, which thermalizes after big bang nucleosynthesis by mixing with the standard model neutrinos. For this, we extend an earlier work which has focused on the background evolution only until the dark sector thermalizes by deriving the full background and perturbation equations of the model and implementing them in an Einstein-Boltzmann solving code. We expound on the behavior of this model, discussing the wide range of parameters that result in interesting and viable cosmologies that dynamically generate dark radiation during a range of epochs. We find that for the strongly self-coupled regime, there is no large cosmological impact for a tight prior on the mass, whereas larger mass ranges allow a smooth interpolation between a behavior close to the $\Lambda$CDM cosmological standard model and close to an additional component of strongly self-interacting dark radiation. In the weakly self-coupled regime we find that we can accommodate a parameter space relevant for the neutrino anomalies as well as one relevant to easing the Hubble tension. 
	\end{abstract}
 
\maketitle

\section{Introduction}\label{sec:intro}

The $\Lambda$CDM standard model of cosmology has proven an excellent descriptive model for most observations in the advancing age of precision cosmology. Similarly, the standard model has remained a valid description of particle physics for many decades now. Despite these successes, there are reasons to believe that neither model is complete. The standard model leaves many questions unanswered, such as the lack of explanation of cosmological observations involving dark matter or dark energy, its connection to quantum gravity, or explanations of the neutrino masses or the strong CP problem. On the other side, the $\Lambda$CDM cosmological model has recently been plagued by a number of tensions, such as the Hubble tension, the significance of which has risen beyond the $5\sigma$ level. This tension persists between the expansion rate as measured locally using a ladder of distance calibrations (such as for example measured by the SH0ES collaboration \cite{Riess:2021jrx}) and the Hubble parameter as indirectly inferred from the CMB anisotropy power spectra (see, for instance, \cite{Planck:2018vyg}; for reviews see \cite{Schoneberg:2021qvd,DiValentino:2021izs,Verde:2023lmm,Freedman:2021ahq,Freedman:2023jcz,Kamionkowski:2022pkx}).

Motivated by the shortcoming of the standard model of particle physics, multiple extensions of its particle content have been invoked which, depending on their nature, can have a strong impact on the CMB anisotropies at the heart of the Hubble tension. Specifically, many extensions predict light degrees of freedom, where the most famous examples are Nambu-Goldstone bosons, chiral femions, and gauge bosons. Such light degrees of freedom behave like a component of dark radiation at the CMB era. These ordinary models with additional dark radiation face problems particularly with predicting overly damped CMB polarization anisotropies at small scales (see for example \cite{Chacko:2015noa, Buen-Abad:2015ova, Chacko:2016kgg,Cyr-Racine:2015ihg,Lesgourgues:2015wza, Brust:2017nmv, Buen-Abad:2017gxg,Archidiacono:2019wdp, Blinov:2020hmc,Allali:2024cji} for assessments of the effects of light species, and \cite{Schoneberg:2021qvd} for a collection of constraints). To accomodate this issue, very recently the Wess-Zumino dark radiation (WZDR) models and their stepped dark radiation generalizations have been proposed \cite{Aloni:2021eaq,Joseph:2022jsf,Schoneberg:2022grr,Allali:2023zbi}. These models circumvent this issue by extending the dark sector by an additional massive particle, whose nonrelativistic transition and subsequent decay heats the dark radiation at a time shortly before or during recombination, thus impacting the larger scales of the CMB anisotropy power spectra more than the smaller scales.

\pagebreak[20]
However, so far no consistent model of thermalization has been presented for these models, which is especially crucial since their abundance should not be generated already at or before big bang nucleosynthesis (BBN), since this would generically disturb the generated light element abundances and thus be tightly constrained by recent observations. Recently Ref.~\cite{Aloni:2023tff} proposed a mechanism to thermalize such stepped dark radiation models after BBN through a neutrino mixing portal that would be expected generically for such models if the particle with the mass step is a sterile neutrino. However, so far the full cosmological impact of such a model involving a thermalizing dark sector with a mass step has not been investigated.

In this work, we conduct a thorough investigation of this model. In particular, we study a model of a dark sector (\DS) which exhibits a mass threshold and consists of a minimum of one massless bosonic species as well as a fermionic dark neutrino species that mixes with the standard model neutrinos ($\nu$). The massless degrees of freedom facilitate interactions in the dark sector, and these interactions decohere the oscillations of the mixed neutrinos. We name this combined model \wzdrnu.

In \cref{sec:mechanism} we set out to derive a consistent set of background and perturbation equations, and we proceed to qualify the impact of the model on cosmology in \cref{sec:impact}. We then obtain the full cosmological constraints from recent data in \cref{sec:results} and we conclude in \cref{sec:conclusions}.

\section{Neutrino Mixing -  Mechanism and cosmological evolution}
\label{sec:mechanism}

\subsection{Mechanism}
The underlying phenomenon of the neutrino portal models that we study is thermalizing the dark sector from neutrino mixing. The mechanism has two essential ingredients: (i) oscillation between the standard model neutrinos and a dark sector fermion (which is a standard model singlet), and (ii) some dark sector interaction to decohere this oscillation. In simplified terms, the SM oscillates to the dark fermion, scatters, and due to this scattering only partially oscillates back, generating a partial abundance of the dark fermion.

In the minimal scenario, the dark sector includes only a single Weyl fermion ($\nu_d$), and the decoherence is due to the weak interactions present for the SM neutrinos (such as $\nu_\alpha + \nu_\beta \to \nu_\alpha + \nu_\beta$ with $\alpha,\beta \in \{e,\mu,\tau\}$). This mechanism was first proposed by~\cite{Dolgov:1980cq,Barbieri:1989ti,Barbieri:1990vx} (see also~\cite{Enqvist:1990ad,Sigl:1993ctk,McKellar:1992ja}), and is famously known as the Dodelson-Widrow mechanism~\cite{Dodelson:1993je} for the production of sterile neutrino dark matter with keV dark fermion masses. %

Unlike Dodelson and Widrow, we are interested in models of dark radiation, and as such we are interested in much lighter dark fermion masses. If the mixing angle is large enough and the dark fermion mass is not too small, the dark sector will reach thermal equilibrium with the standard model. However, this mechanism is not viable to generate dark radiation in a cosmological context. This is because, for the entire parameter space, the dark fermion would fully thermalize prior to big bang nucleosynthesis (BBN), and hence would generate a large abundance of dark radiation during BBN. As was realized very early on~\cite{Barbieri:1989ti}, this is excluded by measurements of the light element abundances arising from BBN (see e.g.~\cite{Schoeneberg2024}).

Recently Ref.~\cite{Aloni:2023tff} showed that in the presence of dark sector self interactions, the thermalization is delayed and generically happens at most a few orders of magnitude above the dark fermion mass, and thus, quite generically, the dark sector can thermalize after BBN. This can be seen from the thermally averaged effective conversion rate of the SM neutrinos into a dark sector fermions which is given by \cite[Eq.~(4)]{Aloni:2023tff}
\begin{equation}
     \langle \Gammaph \rangle = \frac{\frac{1}{4} \sin^2 2\theta_0 \left(3c_\Gamma T_\nu^5 G_F^2 + \alpha_d^2 \frac{T_d^2}{T_\nu}\right)}{\left(\cos 2\theta_0 + \alpha_d \frac{T_d^2}{m_{\nu d^2}} + 18 c_V \frac{G_F^2 T_\nu^5}{m_{\nu d}^2}\right)^2 + \sin^2 2\theta_0} \times e^{-\mdark/T_\nu}~.
     \label{eq: Gamma}
\end{equation}
Here $\theta_0$ is the mixing angle, $\mdark$ is the dark fermion mass, $T_\nu$ is the standard model neutrino temperature, $T_d$ is the dark sector temperature, $G_F$ is the Fermi coupling, $\alpha_d$ is the dark coupling constant, and $c_\Gamma = 7\pi/24 \simeq 0.92$ and $c_V = c_\Gamma \cdot 4 \sin^2(2\theta_W)/(15\alpha) \simeq 23$ 
\cite{ParticleDataGroup:2022pth} (with Weinberg angle $\theta_W$ and fine structure constant $\alpha$) are well known numerical factors~\cite{Dodelson:1993je}. We assume no lepton asymmetry~\cite{Shi:1998km}.

\enlargethispage*{1\baselineskip}
Our expression differs from \cite[Eq.~(4)]{Aloni:2023tff} by the exponent of $-\mdark/T_\nu$ that we introduce for phenomenological reasons. Since massless SM neutrinos with energies below the dark fermion mass are not expected to oscillate into the dark sector neutrinos, we observe that any integration over the momentum of the SM neutrinos distribution function must have a lower bound of $\mdark$\,. To leading order in $\mdark/T_\nu$, this lower bound introduces the exponent which guarantees that the dark sector is not being thermalized once the temperature of the standard model neutrinos drops below $\mdark$\,.\footnote{The exact form of the cutoff is yet to be determined in a future study. However, the polynomial prefactors are not expected to significantly change the results.}

Setting $\Gammaph/H \simeq 1$, we observe that in the relativistic limit $T_\nu \gg \mdark$\,, the thermalization temperature is roughly given by $\Tequil = \mdark \cdot [\theta_0^2 M_{pl}/\mdark]^{1/5}$ (see \cite{Aloni:2023tff}).
While this mechanism is agnostic about the details of the dark sector, for the remainder of this work, we will focus on one minimal realization of it with exactly one dark Weyl fermion ($\nudark$), and one massless real scalar ($\phi$) making up the dark sector.
Moreover, we will take the dark fermion to mix equally with all standard model neutrinos, though we will briefly comment on the case where the dark fermion mixes primarily with only one neutrino.

\subsection{Background evolution}

In order to study the cosmic history of our model, we need to derive the evolution equations for the three species. The Boltzmann equations for the three species ($\nuSM$, $\nudark$, $\phi$) are fully characterized by their interactions as follows
\begin{align}\label{eq:fundamental}
		\frac{\partial f_\nuSM }{\partial \ln a} - p \frac{\partial f_\nuSM}{\partial p}  &= - \frac{\langle \Gammaph \rangle}{H} \left(f_\nuSM - f_{\nu d}\right)~,  \\
		\frac{\partial f_{\nu d} }{\partial \ln a} - p \frac{\partial f_{\nu d}}{\partial p}  &=\frac{g_\nuSM}{g_{\nu d}} \frac{\langle \Gammaph \rangle}{H} \left(f_\nuSM - f_{\nu d}\right) + 
C_{\nudark-\phi}[f_\nudark,f_\phi]
  ~, \\
		\frac{\partial f_\phi }{\partial \ln a} - p \frac{\partial f_\phi}{\partial p}  &= 
-\frac{g_{\nu d}}{g_\phi}C_{\nudark-\phi}[f_\nudark,f_\phi]
+ C_{\phi^n}[f_\phi]~,
\end{align}
where $a$ is the scale factor, $g_i$ is the multiplicity of species $i$, and $f_i$ is its phase-space distribution function with associated momentum $p$.
We explicitly include a $2\leftrightarrow2$ interaction term $C_{\nudark-\phi}$ between $\nudark$ and $\phi$ with a rate which is always larger than $H$, as well as a rapid self-interaction for $\phi$ denoted by $C_{\phi^n}[f_\phi]$. 
We explicitly assume that this self-interaction term together with the additional dark and non-dark interactions can effectively suppress any chemical potential. The prefactors of $g_{\nu d}/g_\phi$ and $g_{\nu}/g_{\nu d}$ come from energy conservation. This altogether implies that the dark fermion and scalar are always thermalized, i.e., $T_\phi = T_{\nu d}$\,.% 

%%%%%%%
In our Boltzmann equations, we explicitly include the thermally averaged $\nuSM\leftrightarrow \nudark$ conversion rate $\langle \Gammaph \rangle$. The conversion rate is in general an energy-dependent process, and hence depends on the distribution functions of the relevant species. However, for the sake of this initial exploration, we focus on the main features of the evolution, which should be captured by this effective treatment. For instance, it is clear that oscillations should become inefficient when the momentum falls below the mass threshold $p<\mdark$\,, and thus we introduced the exponential shut-off of \cref{eq: Gamma}. Deriving the exact equations for oscillations between relativistic and non-relativistic particles is left for future work.
While our simplified treatment of the cutoff mainly affects the tails of the distributions when the temperature drops well below the mass of the dark fermion ($\mdark$), this treatment of averaging over the energy-dependence will require attention to avoid a violation of detailed balance, which we will repeatedly discuss below both for the background and for the perturbations.

%%%%%%%%%%%%%%%%%%%%%
Having the Boltzmann equations in hand, we can now move on to calculate the background evolution of the dark sector and the SM neutrinos. 
We take the dark sector to be initially unpopulated, with $\rho_\phi=\rho_{\nu d}=0$. Moreover, due to the rapid interactions between the dark fermion and scalar, we will treat them as a single interacting fluid with energy density $\rhoDS$\,.
We approximate the standard model neutrinos to be massless, with energy density given by $\rho_\nu = \frac{\pi^2}{30} g_\nu \frac{7}{8} T_\nu^4$ and pressure $3P_\nu = \rho_\nu$ as usual. The initial neutrino temperature ($T_\nu$) is determined such that $\rho_\nu / \rho_\gamma = \frac{7}{8} \left(\frac{4}{11}\right)^{4/3} N_\mathrm{eff}$\,, which gives $(T_\nu/T_\gamma)^4 = \left(\frac{4}{11}\right)^{4/3} [ N_\mathrm{eff}/3 ]$ using $g_\nu=6$ and $g_\gamma =2$ (giving almost the instantaneous decoupling limit of $T_\nu/T_\gamma \approx (4/11)^{1/3}$, but also taking into account the QED and non-instantaneous decoupling contributions to $N_\mathrm{eff}$).

We begin describing the evolution of background quantities by examining the total energy conservation equation $d(\rho a^3) + Pd(a^3) = 0$, which must hold for any closed system. Since the standard model neutrinos stop significantly interacting after the freeze-out of weak interactions at $T_\gamma \simeq 0.7\mathrm{MeV}$, we can treat the combined neutrino+dark sector system as closed in this regime, leaving us with two semi-closed subsystems: the standard model neutrinos ($\nu$), and the dark sector ($\DS$). Therefore we can write down a separate energy conservation equation for each system as long as we take into account and balance the source/sink terms $S$, giving
\begin{align}
	& \frac{\partial \rho_\nuSM}{\partial \ln a} + 4 \rho_\nuSM = \frac{\partial S}{\partial \ln a} ~, \label{eq: rho_nu background}\\
	& \frac{\partial \rho_\DS}{\partial \ln a} + 3 \left(\rho_\DS+ P_\DS\right)= -\frac{\partial S}{\partial \ln a} ~,
\label{eq: rho_ndark background}
\end{align}
where we have conveniently written the equations in terms of derivatives with respect to $\ln a$. We can find the form of the source/sink term by integrating the fundamental Boltzmann equation \cref{eq:fundamental} 
over $\frac{g_\nu}{(2\pi)^3}\int_{0}^{\infty} \mathrm{d}^3p \cdot p$, and matching it with \cref{eq: rho_nu background}. The left hand side of \cref{eq:fundamental} trivially (and by definition) matches the  left hand side of \cref{eq: rho_nu background}. For the right hand side, let us emphasize once again that we assume the conversion rate to be momentum independent, see \cref{eq: Gamma}.
We explicitly integrate over the distribution of the standard model neutrinos, and deduce the pre-factor of the dark sector energy density by imposing detailed balance.  
Note that the precise expression of the source term might thus not be exact, but we expect the form to be correct and corrections to the pre-factor to be subdominant. 
Altogether, we find the following equations for the background quantities (see~\cref{app:background_detailed_balance} for details)
\begin{align}
    & \frac{\partial \rho_\nuSM}{\partial \ln a} + 4\rho_\nuSM  = -\frac{\langle \Gammaph \rangle}{H} \rho_\nuSM\left[1 - \frac{R_{3,\nudark}}{R_{{3,\nudark},{[T \to T_\nu]}}}\right] ~, \label{eq: derivative_density_detailed_balance} \\
    & \frac{\partial \rho_\DS}{\partial \ln a}+ 3\left(\rho_\DS + P_\DS \right) = \frac{\langle \Gammaph \rangle}{H} \rho_\nuSM\left[1 - \frac{R_{3,\nudark}}{R_{{3,\nudark},{[T \to T_\nu]}}}  \right] ~,
\end{align}
which can be re-written in terms of the temperatures of each sector as
\begin{align}
    & \frac{\partial \ln T_\nuSM}{\partial \ln a} + 1  = -\frac{1}{4}\frac{\langle \Gammaph \rangle}{H}\left[1 - \frac{R_{3,\nudark}}{R_{{3,\nudark},{[T \to T_\nu]}}} \right] ~, \label{eq: temperature_evolution_SM}\\
    & \frac{\partial \ln \TDS}{\partial \ln a}+ \frac{\rho_\DS + P_\DS}{\rho_\DS + R_{0,DS}}  =
    \frac{\rho_\nuSM}{3(\rho_\DS + R_{0,DS})}\frac{\langle \Gammaph \rangle}{H} \left[ 1 - \frac{R_{3,\nudark}}{R_{{3,\nudark},{[T \to T_\nu]}}}\right] ~.
\end{align}
As was introduced in~\cite{Schoneberg:2022grr}, we have used $\mathrm{d}\rho_\xi = 3(\rho_\xi+R_{0,\xi}) \mathrm{d}T_\xi$ with the pseudo-density $R_{0,\xi}$. Here, we generalized the pseudo-densities defined in~\cite{Schoneberg:2022grr} to a set of moments of order $n$ for a species labeled by $\xi$ with distribution function $f_\xi$ and $g_\xi$ degrees of freedom to be
\begin{align}
	3R_{n,\, \xi} \equiv \frac{g_\xi}{(2\pi)^3}\int_{0}^{\infty} \mathrm{d}^3p \left[\frac{E^3}{p^2}\right] \left(\frac{p}{E}\right)^n f_\xi (E,T) ~,
\label{eq:pseudo-density}
\end{align} 
where $E$ is the energy and $p$ is the momentum. One can immediately recognize the energy density $\rho_\xi = 3 R_{2,\xi}$ and pressure $P_\xi = R_{4,\xi} $ as well as the pseudo-density of Ref.~\cite{Schoneberg:2022grr} as $R_\xi = R_{0,\xi}$\,. 
In the limit $T_\xi \ll m_\xi$ we find that $R_{n,\, \xi}$ is generically suppressed by the Boltzmann factor $\exp(-m_\xi/T_\xi)$, since $f_\xi \to \exp(-E_\xi/T_\xi)$. 
By $R_{{3,\nudark},{[T \to T_\nu]}}$  we mean that the dark fermion third pseudo-density moment should be evaluated at the standard model neutrino temperature $T_\nu$, as opposed to its own temperature $T_\DS$\,.
The evolution of $\TDS$ and $T_\nuSM$ is shown for $\mdark=1$ eV, $\alphad = 1$, and several choices of $\theta_0$ in \cref{fig:temperature_evolution}.

\subsection{Perturbation Evolution}

We start by deriving the standard model neutrino perturbation equations, which deviate from the usual equations of massless neutrinos (see Ref.~\cite[eq.~(50)]{Ma:1995ey}) by the interaction term. Since the energy conservation of \cref{eq: rho_nu background} is not fulfilled at zeroth order and instead involves a source term, there is an additional term in the equations (which can be derived either from the zeroth order collision term or by a re-interpretation of the Liouville operator, see \cref{app:liouville,app:simplified_liouville}). This extra term takes the form
\begin{equation}
F^\ell_\nuSM \cdot \frac{\mathrm{d}(\rho_\nuSM a^4)}{\mathrm{d}\tau} 
    = F^\ell_\nuSM \cdot a^4\mathcal{H} \frac{\partial S}{\partial \ln a}
    = F_\nuSM \cdot (\rho_\nuSM a^4) \cdot 4 \mathcal{H} \left[\frac{\partial \ln (a T_\nuSM)}{\partial \ln a}\right] ~,\label{eq: Liouvile correction}
\end{equation}
using first \cref{eq: rho_nu background} and then \cref{eq: derivative_density_detailed_balance,eq: temperature_evolution_SM}. As defined below, we use the standard notation for the perturbed quantities $F^\ell_\xi$ (as in e.g. \cite{Ma:1995ey}).
We also have the first order $1\leftrightarrow 1$ collision operator
\begin{align}\label{eq: neutrino fundamental collision term}
    C[\nu] = a\braket{\Gammaph}\left(f^0_\nudark \Psi_\nudark - f^0_\nuSM\Psi_\nuSM\right)
\end{align}
integrated over $\frac{g_\nu}{(2\pi)^3 \rho_\nu} \int \mathrm{d}^3 p \cdot p$. Here we face the same problem as before, namely that this integration is very difficult in terms of the true energy-dependent conversion rate, so we once again approximate by using the thermally averaged rate. Just like with the background quantities, this approximation leads to a violation of detailed balance. As before, we integrate over the distribution function of the standard model neutrinos, and deduce the dark sector counterpart by imposing detailed balance. Following~\cite{Ma:1995ey},
\begin{equation}
    \delta_\nuSM  = \frac{g_\nu}{(2\pi)^3 \rho_\nu} \int \mathrm{d}^3 p \cdot p f^0_\nuSM\Psi_{\nuSM}^0 ~, \quad
    \theta_\nuSM = {\frac{3k}{4}}\frac{g_\nu}{(2\pi)^3 \rho_\nu} \int \mathrm{d}^3 p \cdot p f^0_\nuSM\Psi_{\nuSM}^1 ~, \quad
    F_\nuSM^\ell = \frac{g_\nu}{(2\pi)^3 \rho_\nu} \int \mathrm{d}^3 p \cdot p f^0_\nuSM\Psi_{\nuSM}^\ell ~,
\end{equation}
we find
\begin{align}
    & C_0 [\nuSM] = -a\braket{\Gammaph} \left[\delta_\nuSM-\frac{4}{3}\frac{\delta_\DS}{1 + w_{R_0,\DS}} \right] ~, \\
	& C_1 [\nuSM] = -a\braket{\Gammaph} (\theta_\nuSM - \theta_\DS) ~,\\
    & C_{\ell\geq 2}[\nu] = -a  \braket{\Gammaph} F_\nu^\ell ~.
\end{align}
The simplest terms to understand are the higher moments, $\ell \geq 2$. Since the dark sector is a perfect fluid, all moments with $\ell \geq 2$ vanish. The structure of the dipole is also quite clear, since in the tightly coupled limit one must reach the limit $\theta_\nuSM \to  \theta_\DS$ (moreover, since the dark sector fluid is tightly coupled, we always have $\theta_\nudark=\theta_\phi=\theta_\DS$). Finally, for the monopole, it is important to realize that in the limit of local thermal equilibrium, $\delta T_\DS = \delta T_\nu$\,.
By using $\mathrm{d}\rho/d\mathrm{T} = (\rho+R_0)$ and also using the adiabatic approximation, we find the relation between $\delta_\DS$ and $\delta_\nu$. A more detailed derivation can be found in \cref{app: monopole vs temperature}.

% %
For the interacting standard model neutrinos, we follow the infinite hierarchy of multipoles, since they decouple and start to free steam at late times.
The {\it standard model neutrino ($\nuSM$) perturbations} are given by:
\begin{equation}
    \begin{split}
        \frac{\partial \delta_\nuSM}{\partial \tau} \,+\, \frac{4}{3}\theta_\nu &\,-\, 4 \dot{\phi}_{CN}   \\ &
		= - 4 \mathcal{H} \left[\frac{\partial \ln (a T_\nuSM)}{\partial \ln a}\right] \delta_\nuSM 
		-a\braket{\Gammaph}
   \left[\delta_\nuSM-\frac{4\delta_\DS}{3(1 + w_{R_0,DS})} \right] ~,
    \end{split}
    \label{eq: delta_nu equation} 
\end{equation}
\begin{equation}
    \begin{split}
    \frac{\partial \theta_\nuSM}{\partial \tau} -k^2 \left[\frac{1}{4}\delta_\nuSM -\sigma_\nuSM \right] + k^2\, \psi_{CN}
		=- 4\mathcal{H}\left[\frac{\partial \ln (a T_\nuSM)}{\partial \ln a}\right]\theta_\nuSM 
		+ a\braket{\Gammaph}\left(\theta_\DS-\theta_\nuSM\right) ~,
    \end{split}
    \label{eq: theta_nu equation} 
\end{equation}

\begin{equation}
    \begin{split}
    \frac{\partial F^\ell_\nu}{\partial \tau}  - \frac{k}{2\ell+1} \left[\ell\,F^{\ell-1}_\nu - (\ell+1)F_\nuSM^{\ell+1}\right]
		= - 4  \mathcal{H}\left[\frac{\partial \ln (a T_\nuSM)}{\partial \ln a}\right]
 F^\ell_\nu   - a\braket{\Gammaph}F^\ell_\nuSM \quad , ~ \ell\geq2
		~,
    \end{split}
    \label{eq: Fl_nu equation} 
\end{equation}
where $\phi_{CN}$ and $\psi_{CN}$ are the metric potentials in the conformal Newtonian gauge, $\tau$ is the conformal time, $w = P/\rho$ is the equation of state, and $w_{R_0,x} \equiv R_{0,x}/\rho_x$ is the pseudo equation of state.

To obtain the analogous perturbation equations for the dark sector, we make two important choices. First, as before, we treat the dark sector as perfectly coupled, allowing us to treat it as a single interacting fluid, which also allows us to set $F^{\ell\geq2}_{\nudark} = F^{\ell\geq2}_\DS=0$. Second, we use the fact that the combination of dark sector and standard model neutrinos is overall conserved to find the corresponding collision terms. 
Note that we naturally find factors of $-\rho_\DS/\rho_\nuSM$ appearing from the energy conservation (and $-(\rho_\DS+P_\DS)/(\rho_\nuSM+P_\nuSM)$ from the momentum conservation). More details can be found in \cref{app:simplified_liouville}.

Ultimately, we find the {\it dark sector (DS) perturbations}:
\begin{equation}
    \begin{split}
        \frac{\partial\delta_\DS}{\partial \tau} & \,+\, (1+w_\DS)\left(\theta_\DS \,-\, 3\dot{\phi}_{CN}\right) 
 \,+\, 3\mathcal{H}\left(c_{s,DS}^2-w_\DS\right)\delta_\DS  \\ 
        & = 4 \mathcal{H} \left[\frac{\partial \ln (a T_\nuSM)}{\partial \ln a} \frac{\rho_\nuSM}{\rho_\DS}\right]\delta_\DS
        \,+\, a\braket{\Gammaph}\frac{\rho_\nu}{\rho_\DS} \left[\delta_\nuSM  - \frac{4\delta_\DS}{3(1 + w_{R_0,DS})}\right]~,
    \end{split}
    \label{eq: delta_DS equation} 
\end{equation}
\begin{equation}
    \begin{split}
     \frac{\partial\theta_\DS}{\partial \tau} \,-\, k^2 \frac{ c_{s,DS}^2}{1+w_\DS}&\delta_\DS \,+\, \mathcal{H}\left(1-3c_{s,DS}^2\right)\theta_\DS \,+\, k^2 \psi_{CN} \\
	& = 
        4\mathcal{H} \left[\frac{\partial \ln (a T_\nuSM)}{\partial \ln a}\frac{\rho_\nuSM}{\rho_\DS}\right]\theta_\DS
        \,+\, a\braket{\Gammaph} \left[\frac{4\rho_{\nuSM}}{3(\rho_\DS+P_\DS)}\right]\left(\theta_\nuSM - \theta_\DS\right)~. 
    \end{split}
    \label{eq: theta_DS equation} 
\end{equation}

%%%%%%%%%%%%%%%%%%%%%%%%%%%%%%%%%%%%%%%%%
%%%%%%%%%%%%%%%%%%%%%%%%%%%%%%%%%%%%%%%%%
%%%%%%%%%%%%%%%%%%%%%%%%%%%%%%%%%%%%%%%%%
\section{Qualitative impact of model parameters}\label{sec:impact}
\enlargethispage*{4\baselineskip}
Before showing the results of the Bayesian inference, we build intuition about the model behavior by exploring the phenomenology and cosmological impact of the new model parameters $\{\theta_0,\, \alphad,\,\mdark\}$. Given that similar models with mass thresholds in the dark sector causing a step in their abundance have been shown to display a decent ability to ease the Hubble tension (see for example the Wess-Zumino dark radiation model (WZDR), and generalizations \cite{Aloni:2021eaq,Joseph:2022jsf,Schoneberg:2022grr,Allali:2023zbi}), we focus on a similar regime of parameter space in this work.
In particular, we are interested in choices of $\theta_0$\,, $\alphad$\,, and $m_\nudark$ which result in a dark sector which is thermalized after BBN ($m_\nudark < T_\mathrm{BBN}\sim 100$ keV) in order not to disturb light element abundances, and a mass threshold which can be relevant for recombination (focusing on $m_\nudark \sim \mathcal{O}(0.1)-\mathcal{O}(10)$ eV) in view of the Hubble tension. From this point, we will refer to the model we study as the \wzdrnu model.

\begin{figure}
    \centering
    \includegraphics[width=0.5\textwidth]{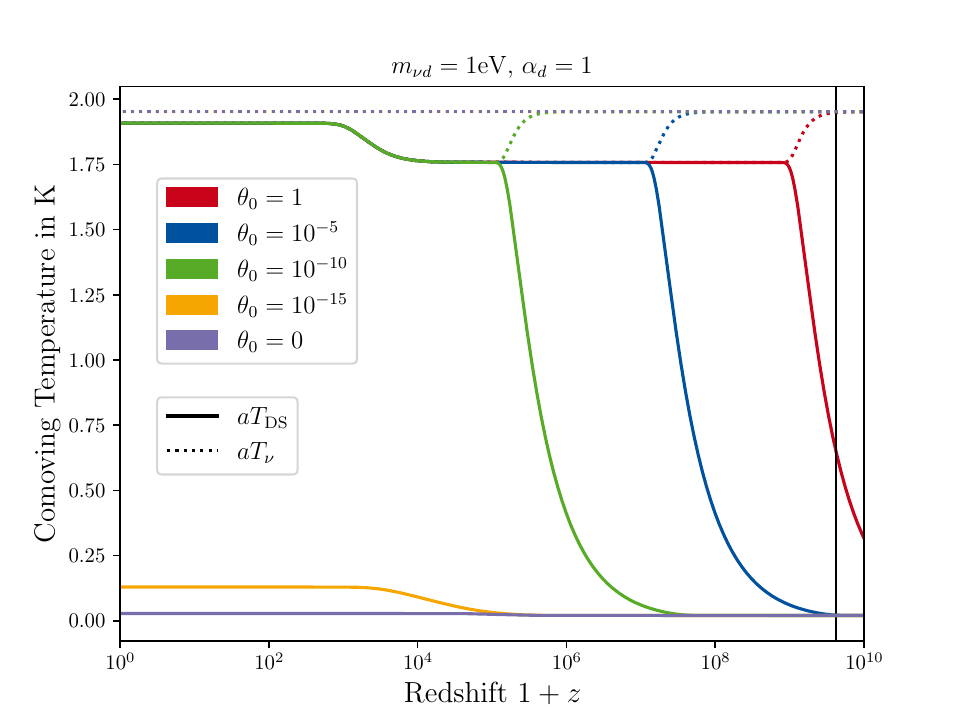}
    \caption{A typical evolution of the dark sector and standard model neutrino temperatures in the model. We show the comoving temperatures, which are simply defined as $a \cdot T$. The vertical line shows the approximate redshift corresponding to the photon thermal bath reaching a temperature of $1 \, \mathrm{MeV}$, which roughly corresponds to a time around which big bang nucleosynthesis occurs.}
    \label{fig:temperature_evolution}
\end{figure}

\begin{figure}
    \centering
    \includegraphics[width=0.5\textwidth]{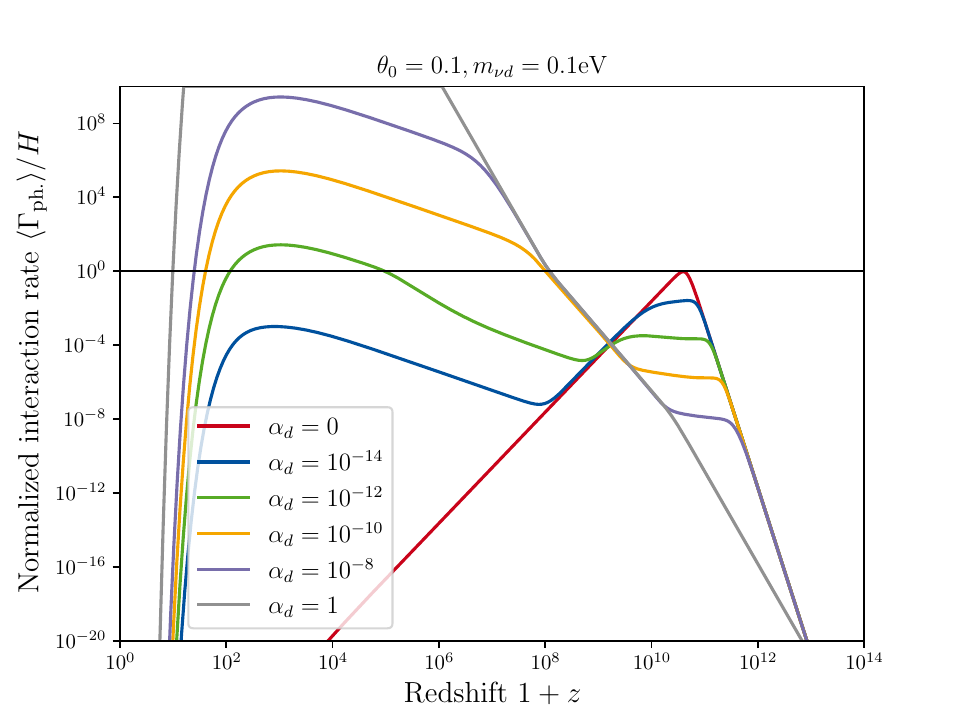}
    \caption{The evolution of the interaction rate $\langle \Gammaph \rangle$ as a function of redshift for various values of the $\alpha_d$ dark coupling constant. We fix the mixing angle $\theta_0$ and mass $m_\nudark$ to some exemplative values.}
    \label{fig:alpha_dependence}
\end{figure}

First, we show in \cref{fig:temperature_evolution} a typical evolution of the \wzdrnu model with $\alpha_d = 1$ for the dark sector and standard model neutrino temperature (the strongly coupled case we investigate in \cref{ssec:strongly_coupled}). We can quickly appreciate that for large values of $\theta_0$ the standard model and dark sectors thermally couple earlier, while for lower values of $\theta_0$ the sectors couple later. We also observe the step-like rise in temperature at a redshift corresponding to the dark fermion mass (in the figure it is $\mdark = 1 \mathrm{eV} \to z_\mathrm{step} \sim 6600 \simeq 10^{3.8}$) due to the non-relativistic transition of the dark fermion, the decays of which heat the remaining dark sector. 
If the value of $\theta_0$ is small enough such that the non-relativistic transition of the dark fermion occurs before the two sectors would in principle couple, no such coupling can occur as the interaction rate becomes Boltzmann suppressed before it rises enough from the temperature dependence (purple line in \cref{fig:temperature_evolution}). In this limit, we recover the $\Lambda$CDM model approximately.

Next we explore the role of the dark coupling constant $\alphad$\,. A priori there is a large degeneracy between $\alphad$ and $\theta_0$\,, as both cause an overall rescaling of the interaction rate~$\langle \Gammaph \rangle$ of \cref{eq: Gamma}. However, in the limit of very small $\alphad$, one begins to recover the Dodelson-Widrow mechanism with an early thermalization and then immediate decoupling. In \cref{fig:alpha_dependence} we show the thermally averaged conversion rate $\langle\Gammaph\rangle$ normalized to the Hubble expansion rate of the universe. 

\pagebreak[5]
When $\alphad=0$ (red curve), we only have the Dodelson-Widrow peak of an early thermalization at $z \simeq 10^{11}$ in this case. As we increase the value of $\alphad$, the initial peak is suppressed due to the additional interactions, while a second continuously growing part is added. If this growing part becomes the dominant contribution and the Dodelson-Widrow peak is entirely suppressed, we recover the phenomenology of the strong-coupling regime discussed above (in~\cref{fig:temperature_evolution}) and later in \cref{ssec:strongly_coupled}. Note that we also observe the exponential suppression for $T_\nu \ll m_{\nu d}$ (which corresponds to $z \sim 100$ in this case).
We also observe that interesting new cases open up in the weakly-interacting regime which are shown in \cref{fig:alpha_dependence} (for example the green line). At early times this case does not thermalize fully, and then subsequently re-couples at a later time. We investigate such weakly-coupled cases in \cref{ssec:weakly_coupled}.

\begin{figure}
    \centering
    \includegraphics[width=0.45\textwidth]{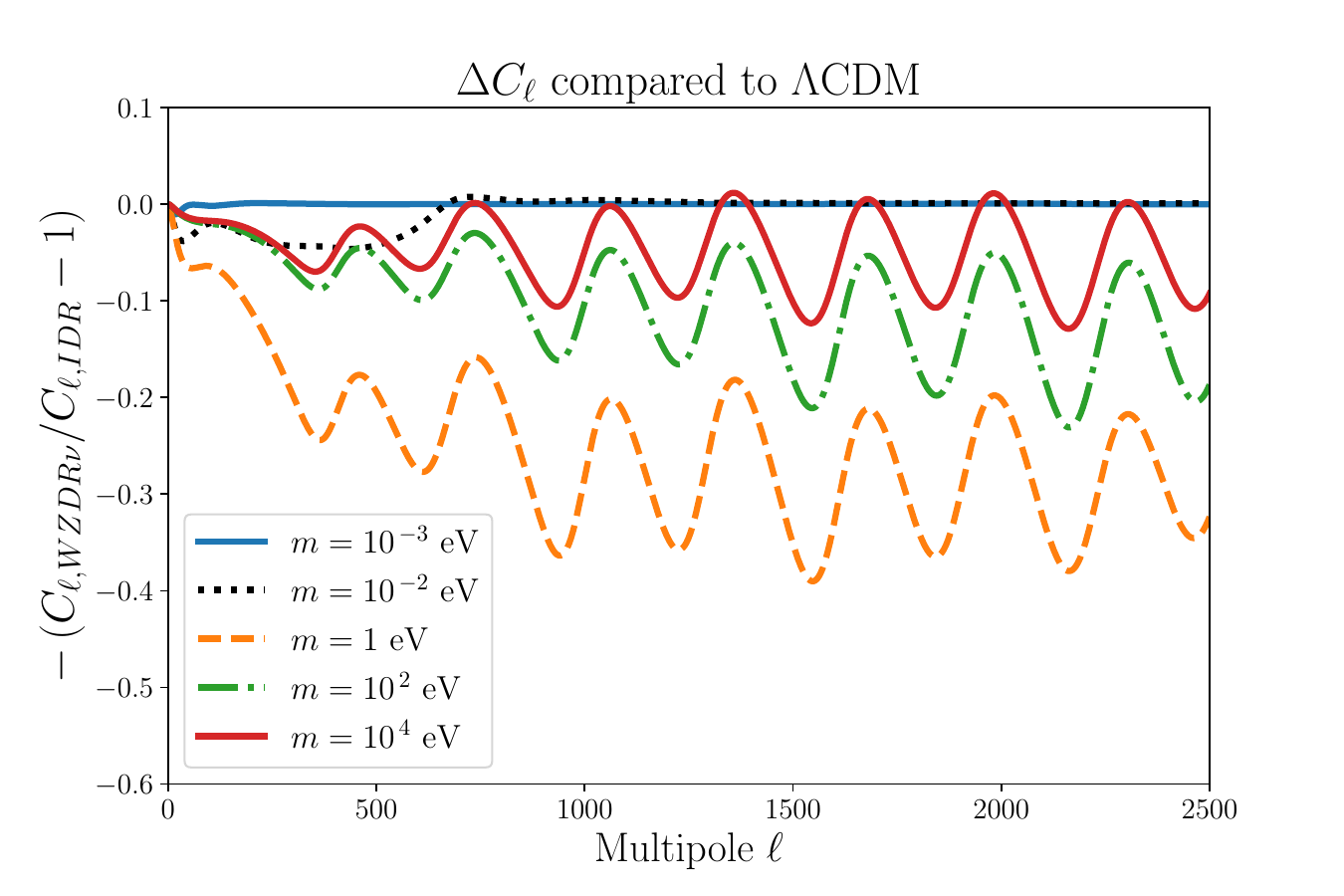}
    \includegraphics[width=0.45\textwidth]{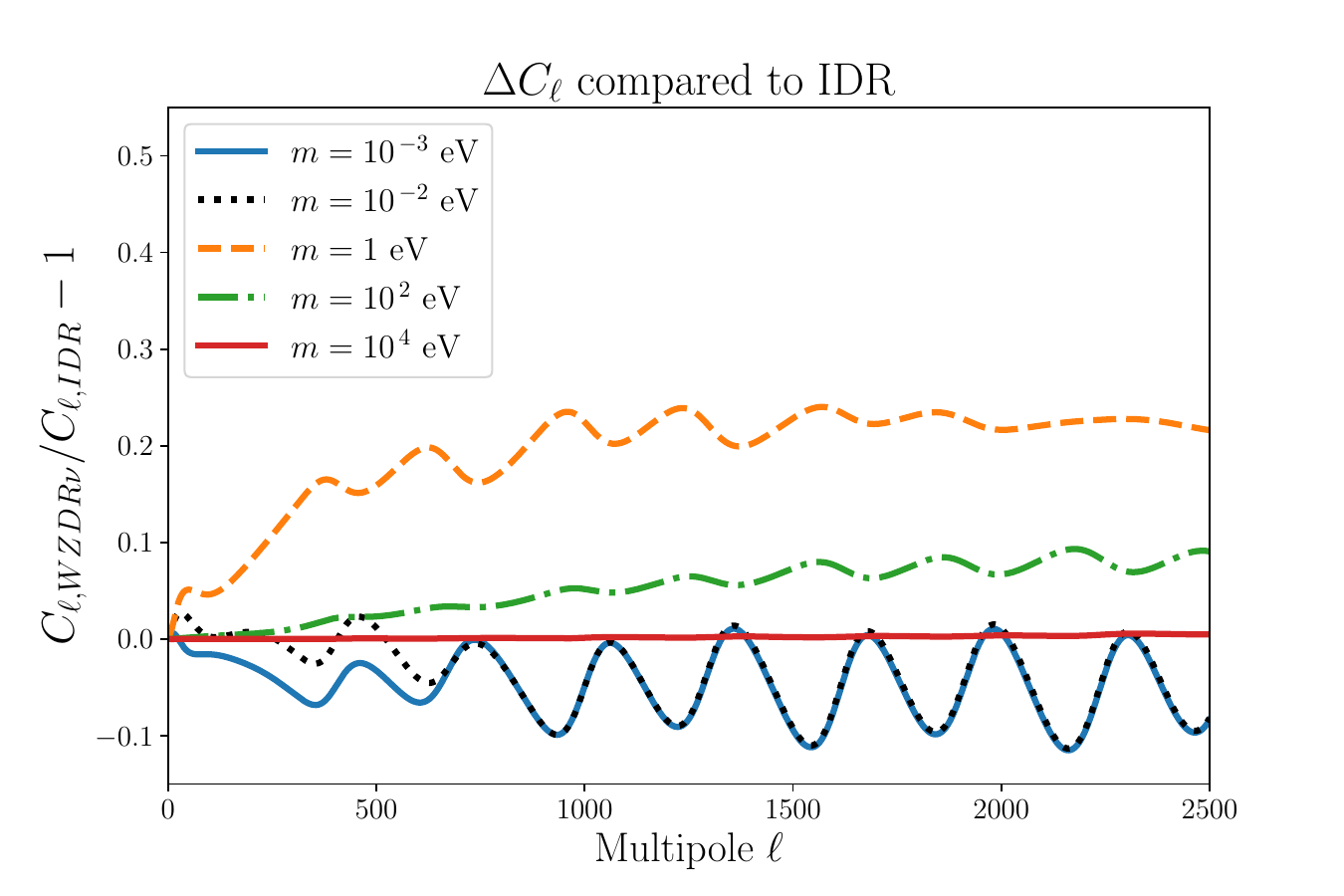}
    \caption{\label{fig:c_ell}Plot of the fractional change in the temperature auto-correlation CMB power spectrum in the \wzdrnu model for various masses ($\mdark$). The parameters of the model to be compared with are adopted to reproduce the expected limits as discussed in the text. Note that the right panel shows the negative difference to facilitate the comparison between the red line in the left panel and the blue line in the right panel. Left: Change with respect to $\Lambda$CDM. Right: Change with respect to a model with self interacting dark radiation (IDR) on top of $\Lambda$CDM. %
    }
    
\end{figure}

Finally, we explore the role of the dark fermion mass $\mdark$ with a focus on its impact on the CMB temperature anisotropy power spectrum. While this scale always directly guides the redshift of decoupling and of the mass step, in the limit of strong coupling ($\alphad \to 1$) the mass also controls the time of thermalization. This is because in the absence of an initial Dodelson-Widrow peak, the temperature of thermalization is proportional to $({\mdark})^{4/5}$ (see \cite{Aloni:2023tff}).

There are two interesting limits to explore. In one limit of high dark fermion mass $\mdark$\,, the dark sector decouples from the standard model neutrinos already before the highest wavenumber relevant for current CMB experiments enters the Hubble horizon. In this limit, the model has the same impact on the CMB anisotropies as one with the same amount of free-streaming/self-interacting (i.e., perfect fluid) dark radiation abundances as the standard model neutrino/dark sector abundances. In the other limit of small dark fermion mass, the dark sector begins to couple only after the time of recombination, and thus, there is no impact of the model before recombination and the impact on the CMB can be considered equivalent to $\Lambda$CDM.

For the rest of this section, we work in the strong coupling regime $\alphad=1$, and we fix $\theta_0=10^{-10}$ such that the dark sector thermalizes after BBN for the range of masses we consider. We show in \cref{fig:c_ell} a comparison of this case to a $\Lambda$CDM model and a \ffs model for the CMB temperature autocorrelation multipole coefficients ($C_\ell$). The free-streaming (FS) plus self-interacting (IDR) model has $\Niur \simeq 2.8$ and $\Nfld \simeq 0.5$, which correspond to the low-redshift abundances of the chosen \wzdrnu model. We immediately notice that in the low mass regime, the $\Lambda$CDM angular power spectra are recovered (blue line, left panel), while in the high mass regime the \ffs power spectra are recovered (red line, right panel), validating to some extent our model in these limits.

For intermediate masses (around $1 \mathrm{eV}$), the \wzdrnu model is tightly coupled exactly around recombination, thus showing the highest deviations from both \LCDM and the mixed \ffs models. Towards smaller masses, the model begins to resemble \LCDM except for a small bump at low multipoles, which enter the Hubble horizon at a later time and thus are still partially sensitive to the \wzdrnu thermalization even for smaller masses. Instead, towards larger masses, the model begins to resemble the \ffs model except for a trend of growing deviation for larger multipoles which have entered earlier and thus experienced more of the \wzdrnu thermalization process. Indeed, in this part of the parameter space the \wzdrnu model gains a phase-shift of the acoustic oscillations imprinted in the CMB.

\section{Results}\label{sec:results}

After building our intuition in \cref{sec:impact}, we present the Bayesian constraints on the \wzdrnu model in the context of cosmological data, exploring two different limits: the strong coupling regime ($\alphad=1$), and the weak coupling regime ($\alphad\ll1$).

The background and perturbation evolution of the \wzdrnu model is computed using a modified version of {\tt CLASS}~\cite{Lesgourgues:2011re,Blas:2011rf}. We performed a suite of Markov-chain Monte Carlo (MCMC) analyses, making use of the {\tt MontePython}\footnote{Available at \href{https://github.com/brinckmann/montepython_public}{\tt https://github.com/brinckmann/montepython\_public}.}~\cite{Audren:2012wb, Brinckmann:2018cvx} sampler, to compare various realizations of the model to several datasets, outlined below. In each case, we assume flat priors for the six parameters of the $\Lambda$CDM model $\{\Omega_b h^2$, $\Omega_\text{cdm} h^2$, $H_0$, $\ln(10^{10} A_s)$, $n_s$, $\tau_\text{reio}\}$. Note that neutrinos, whether interacting or free-streaming, are assumed to be massless. In addition, we study various choices for the free parameters of the \wzdrnu model; whenever left to vary, priors on $\theta_0$, $\mdark$ (corresponding to $z_t$), and $\alphad$ are logarithmic (unless otherwise indicated), allowing us to check many orders of magnitude of parameter space for viability.

We perform most analyses with the following combination of data, referred to as the \enquote{baseline} dataset in this work:
\begin{itemize}
     \item {\bf Planck}: The Planck 2018 high-$\ell$ and low-$\ell$ temperature and polarization (TT, TE, EE) power spectra and lensing reconstruction data~\cite{Aghanim:2019ame}
     \item {\bf BAO}: Baryon acoustic oscillations (BAO) measurements from 6dFGS at $z = 0.106$~\cite{Beutler:2011hx}, SDSS MGS at $z = 0.15$~\cite{Ross:2014qpa}, and CMASS and LOWZ galaxy samples of BOSS DR12 at $z = 0.38$, $0.51$, and $0.61$~\cite{Alam:2016hwk}
     \item {\bf Pantheon}: The Pantheon$+$ catalog of type Ia supernovae between $z=0.001$ and $z=2.26$ \cite{Brout:2022vxf}.
\end{itemize}
Fitting to this baseline dataset will serve our main purposes to provide comparisons between the \wzdrnu model and $\Lambda$CDM. In addition, for the purposes of assessing the viability of resolving the $H_0$ tension, we will also add the following additional prior:
\begin{itemize}
     \item {\bf $+$SH0ES}: The combinaton of {\bf Planck} and {\bf BAO} as outlined above, along with the combined type Ia supernovae data from Pantheon$+$ and SH$0$ES \cite{Riess:2021jrx} as outlined in \cite{Brout:2022vxf} (ranging from $z=0.15$ to $z=2.26$).\footnote{We note in passing that the combined Pantheon$+$SH$0$ES likelihood from \url{https://pantheonplussh0es.github.io/} released in \texttt{MontePython} typically has a lower chi-square than the Pantheon$+$ standalone likelihood due to the restricted lower redshift limit.}
 \end{itemize}

The plots are produced with \texttt{liquidcosmo}\footnote{Available at \url{https://github.com/schoeneberg/liquidcosmo}.} (which is using \texttt{getdist} \cite{Lewis:2019xzd}). 
\Cref{tab:chisquares} summarizes the best fitting chisquare values, while additional information is supplied in \cref{app:tables}, in which \cref{tab:posteriors} gives the posterior statistics (such as mean and uncertainty) for a selection of model and cosmological parameters for each analysis and \cref{tab:bestfits} summarizes the corresponding bestfits.

\subsection{Strongly coupled}\label{ssec:strongly_coupled}

  \begin{figure}
     \centering
     \includegraphics[width=0.7\textwidth]{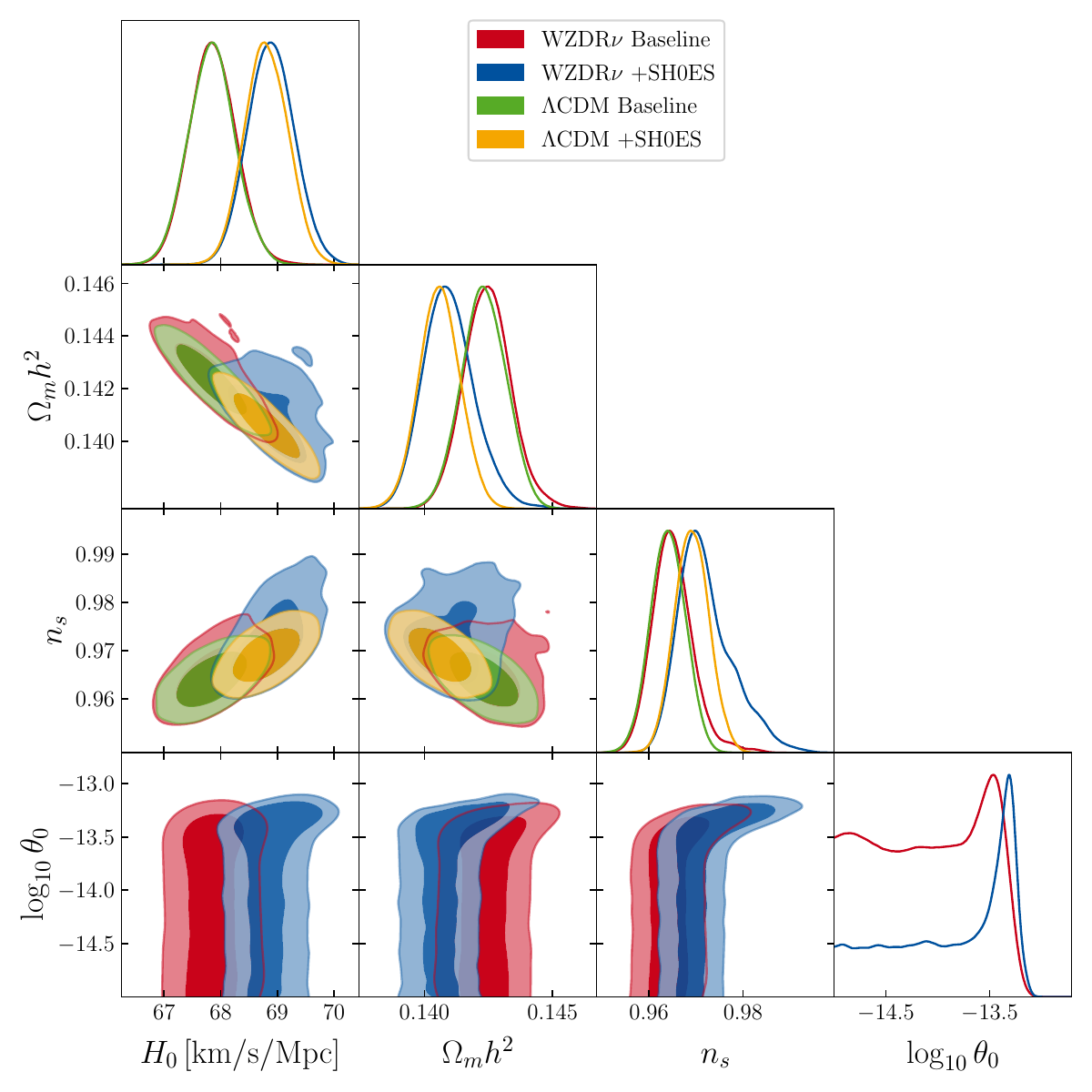}
     \caption{\label{fig:results_baseline} One-dimensional posteriors and two-dimensional constraints at 68\% and 95\% CL for a range of parameters. Shown are the $\Lambda$CDM model and \wzdrnu model (with $\alphad=1$) using the baseline and baseline+SH0ES datasets. Here, we take the mass $\mdark \sim \mathcal{O}(\mbox{eV})$, motivated by dark radiation models which have a mass-threshold slightly before recombination.}
 \end{figure}
 
 \begin{figure}
     \centering
     \includegraphics[width=0.7\textwidth]{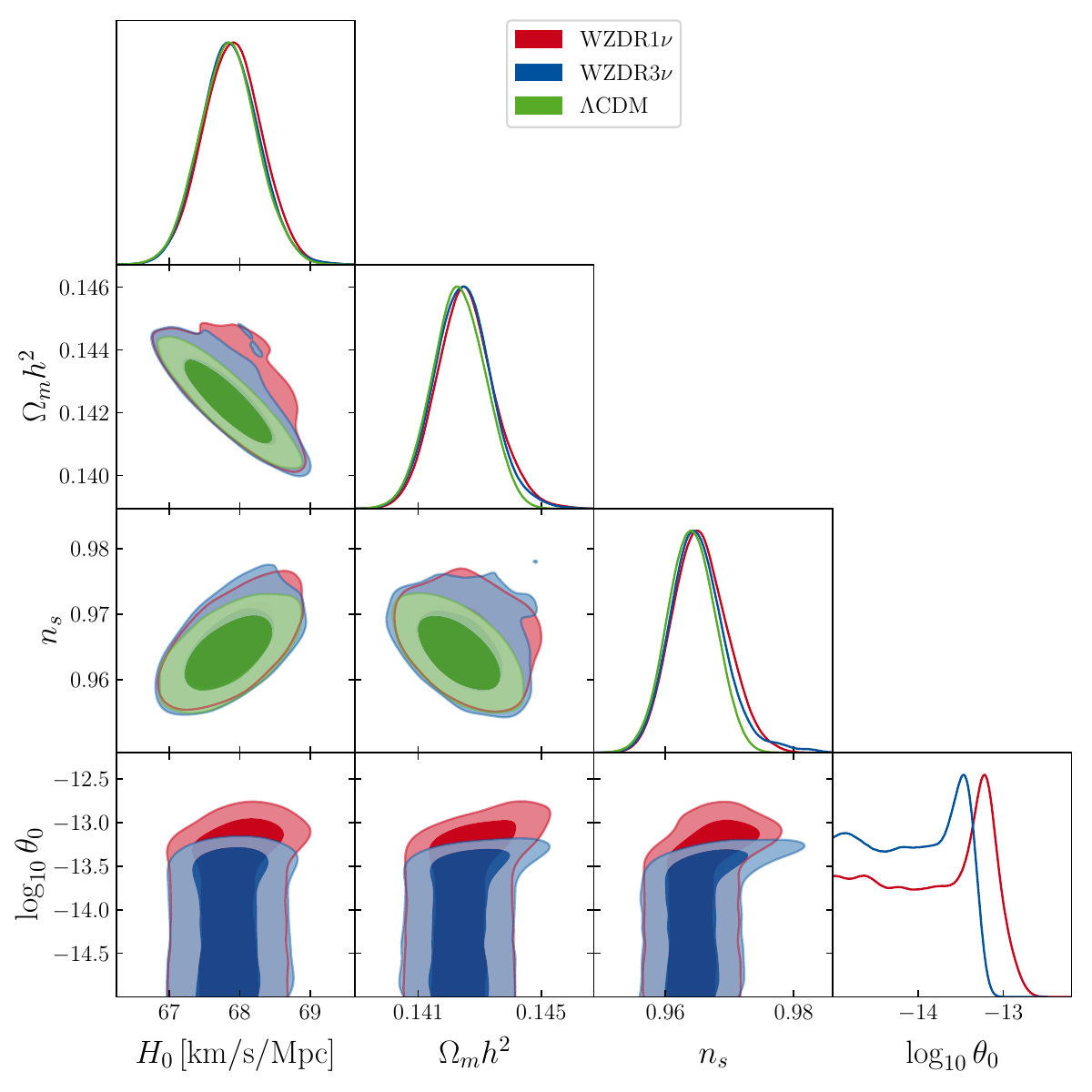}
     \caption{\label{fig:1nu}  One-dimensional posteriors and two-dimensional constraints at 68\% and 95\% CL for a range of parameters. Shown are the $\Lambda$CDM model and two \wzdrnu models, with $\alphad=1$ and either only one or all three of the standard model neutrinos interacting, using the baseline dataset. Here, we take the mass $\mdark \sim \mathcal{O}(\mbox{eV})$, motivated by dark radiation models which have a mass-threshold at this scale.}
 \end{figure}
 
In this section, we investigate the model in the limit in which the dark coupling constant $\alpha_d$ of \cref{eq: Gamma} is large. For concreteness, we set $\alpha_d = 1$, but all order-unity values give approximately the same results. Motivated by the $z_t$ prior of \cite{Aloni:2021eaq,Schoneberg:2022grr,Allali:2023zbi}, we first investigate the case of logarithmic prior of masses in the range $0.1\mathrm{eV}-10\mathrm{eV}$.
We show the constraints on the strongly coupled \wzdrnu model from the baseline dataset as well as the additional prior on the Hubble constant in \cref{fig:results_baseline}, but we marginalize over and do not show the mass explicitly as it does not display any interesting correlations. 
We find that the limit towards small $\theta_0 \lesssim 10^{-14}$ corresponding to the decoupled case gives largely the same constraints on all cosmological parameters as in $\Lambda$CDM, as expected. Instead, the limit of larger $\theta_0$\,, where the dark sector is strongly coupled to the standard model neutrinos, is largely excluded by the data. Indeed, only minor deviations from the $\Lambda$CDM parameters are compatible with the data, mostly in the scalar tilt $n_s$ and the physical matter density $\omega_\mathrm{m} = \Omega_\mathrm{m} h^2$. The impact of the model on the Hubble parameter is somewhat minor, even when an additional prior $+$SH0ES is imposed. The same remains somewhat true when only one of the standard model neutrinos is interacting with the dark sector, as we show in \cref{fig:1nu}. However, in this case, slightly larger couplings are allowed as fewer neutrinos are interacting. The comparison of the minimal chi-square $\chi_\mathrm{eff}^2 = - 2 \ln \mathcal{L}$ from \cref{tab:chisquares} in these cases reveals that these cases do not provide significantly better fits to the data compared to $\Lambda$CDM.

\pagebreak[20]
At first, it might seem counter-intuitive that such a large part of the parameter space with high $\theta_0$ is excluded, even though the presence of a component of strongly interacting dark radiation itself (without coupling to standard model neutrinos) is even slightly preferred by the CMB if it makes up around 10\% of the dark radiation \cite{Blinov+2020} (see also \cite{Allali:2024cji} for the effects of BAO). As such, purely from the perspective of the background abundances, we would not expect this model to be disfavored. Instead, we attribute this strong dislike of the CMB for such strongly coupled models to the large additional impact on the perturbations of the standard model neutrinos, which is generated from \cref{eq: delta_nu equation,eq: theta_nu equation,eq: Fl_nu equation} and visible in the right panel of \cref{fig:c_ell}. As long as the interaction rate is larger than the Hubble rate (at any point before the mass step), this impact is large enough to shift the photon density and shear oscillations, which is observable in the CMB anisotropies.

Nevertheless, it is not trivial that this impact of the interactions on the perturbations is harmful, as neutrino interactions were studied in the context of self-interactions~\cite{Cyr-Racine:2013jua,Oldengott:2014qra,Cyr-Racine:2015ihg,Oldengott:2017fhy,Kreisch:2019yzn,Das:2020xke}, as well as the Majoron model~\cite{{Escudero:2019gvw,EscuderoAbenza:2020egd,Escudero:2021rfi,Sandner:2023ptm}}, and were found to not be completely disfavored by the data.
One difference that we can readily point out is that these
models do not have an interaction at the level of the neutrino velocity or overdensity, which are present in our case.

\begin{figure}
    \centering
    \includegraphics[width=0.45\textwidth]{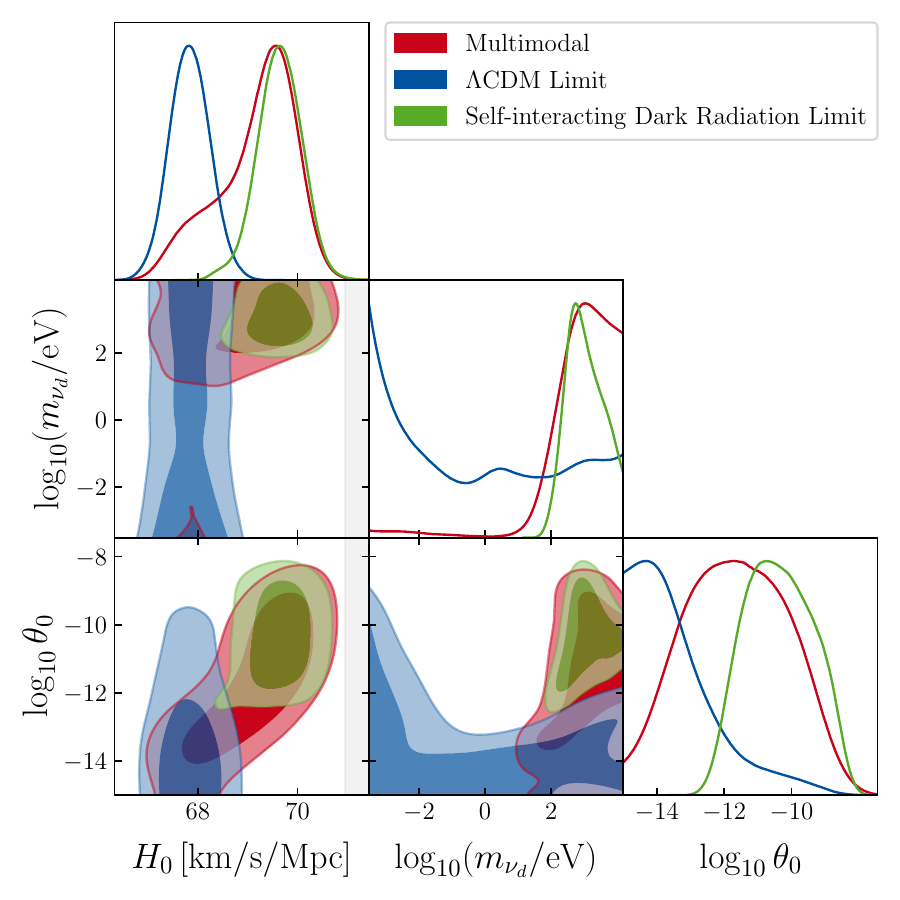}
    \includegraphics[width=0.45\textwidth]{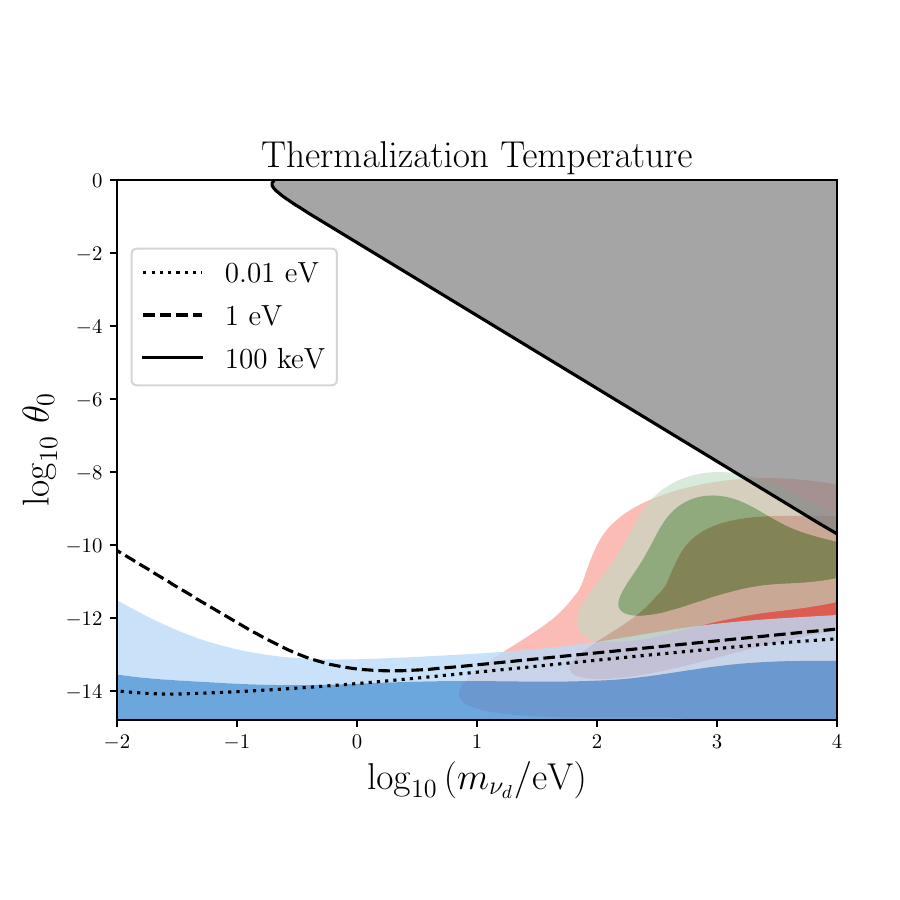}
    \caption{Left: One-dimensional posteriors and two-dimensional constraints at 68\% and 95\% CL for $H_0$, $\mdark$, and $\theta_0$ using the baseline dataset. The constraints shown are derived from three sets of MCMC chains with identical priors on \wzdrnu model parameters. These three cases differ by whether the MCMC explores first the $\Lambda$CDM limit of the model (small $\theta_0$, in blue), or the self-interacting dark radiation limit (large $\mdark$, large $\theta_0$, in red and green). The green shows the constraints before the MCMC locates a better fit in the $\Lambda$CDM limit, while the red shows after the MCMC begins to approach this regime in parameter space. The 95\% confidence interval for the SH0ES measurement of $H_0$ is shown by the light gray shaded region. Right: Thermalization temperature as a function of the mass $\mdark$ and interaction rate parameter $\theta_0$. Constraints on these two parameters are shown, as in the left-side figure. The dark gray region corresponds to thermalization during or before BBN.}
    \label{fig:multimodal_thermalization}
\end{figure}

Given that we recover the $\Lambda$CDM and \ffs limits in the low- and high-mass regimes as seen in \cref{fig:c_ell}, we could expect that allowing a much wider range of dark fermion masses (and corresponding redshifts $z_t$ of the mass step and decoupling) may alleviate these constraints.
This region also covers a significant regime of the parameter space that was proposed as viable for cosmology by~\cite{Aloni:2023tff}.
Exploring this wide range of masses, however, turns out to be challenging. Both the low-mass limit in which the \wzdrnu model resembles $\Lambda$CDM and the high-mass limit in which the model resembles \ffs dark radiation can be efficiently explored by the MCMC algorithm. These cases are shown in blue and green contours in \cref{fig:multimodal_thermalization} (a comparison of the broader and tighter prior range directly is shown in \cref{fig:extended_range}). However, the transition from one regime to the other crosses through a range of intermediately coupled models. For these intermediately coupled models the cosmological parameters such as $h$ and $\Omega_m h^2$ need to rapidly change in order to account for the rapid change in dark sector abundances (see \cref{fig:temperature_evolution}). This means that the intermediate region is hard to cross from the perspective of an MCMC algorithm. Explicitly, looking at \cref{tab:chisquares} we can see that the minimal effective chi-squares of these cases are higher than that of the low-mass case with the $\Lambda$CDM limit (blue contour). Indeed, the green case shows the highest chi-square difference ($+3.1$ to $\Lambda$CDM), the red an intermediate chi-square difference ($+1.0$ to $\Lambda$CDM), and the base case a slight improvement ($-0.6$ to $\Lambda$CDM). In principle this would be resolved using nested sampling algorithms, though their comparatively longer runtime has prevented efficient exploration so far -- we leave exploration of the full posterior with novel sampling tools to future work. Instead we show in \cref{fig:multimodal_thermalization} a red contour that is derived from pushing the chain in the \ffs case with a much higher number of points and new covariance matrices for the sampler. It is, however, worth mentioning that even for this case one would expect the parameter space to be more fully explored with even more runtime. 

The right panel of \cref{fig:multimodal_thermalization} shows the two-dimensional posterior contours for the mass and mixing angle compared to lines indicating the temperature at which a fully coupled model ($\alphad=1$) thermalizes as a function of the same two parameters. The gray region indicates models which thermalize before $T\sim 100 \,\mathrm{keV}$, which we explicitly exclude in our runs in order to avoid larger abundances of light species at BBN. We observe the same behavior as in \cite{Aloni:2023tff}. The constraints are mostly such that the model couples after the CMB ($T \simeq 1\mathrm{eV}$), thus leaving the CMB anisotropies mostly the same (blue contours). However, since the \ffs limit also presents decent fits in terms of CMB anisotropies, the range of $m \gtrsim 10\mathrm{eV}$ (which behaves very similarly to the \ffs case) also exhibits higher couplings $\theta_0$ that already couple and decouple from the standard model neutrinos before the CMB.

We note that in all of these cases, due to the order-unity coupling $\alpha_d=1$, there is a direct relation between the thermalization and the redshift at which $\langle \Gamma \rangle / H > 1$ starts to be fulfilled. As such, it is very hard to find a regime where the dark sector thermalizes without causing problems with the CMB anisotropies, as discussed above (requiring very large masses, which cause the \wzdrnu model to become essentially equivalent to a \ffs model). In \cref{ssec:weakly_coupled} below, we show how releasing the assumption of $\alphad \simeq 1$ can give cosmologically more interesting results.

\subsection{Weakly coupled}\label{ssec:weakly_coupled}

In this section we explore the weak coupling limit $\alphad \ll 1$. As mentioned before, as far as cosmology is concerned, this parameter is somewhat degenerate with the mixing angle $\theta_0$. While it is interesting by itself to explore the different dependence of these two parameters on the cosmology, the regime of weak coupling also allows for much higher mixing angles which are motivated by the neutrino anomalies (see e.g.~\cite{Dasgupta:2021ies} for a review).

It was early realized~\cite{Barbieri:1989ti,Barbieri:1990vx} that sterile neutrinos with masses at the eV scale and large mixing angles, such as favored by the neutrino anomalies, thermalize prior to BBN and hence are incompatible with cosmological bounds on $\Neff$ (see~\cite{Hamann:2011ge} for more recent analysis). 
Refs.~\cite{{Hannestad:2013ana,Dasgupta:2013zpn}} were the first to show that thermalization of sterile neutrinos can be delayed or avoided by introducing sterile neutrino interactions, and the mechanism was further investigated by many follow-up works~\cite{Mirizzi:2014ama,Tang:2014yla,Chu:2015ipa,Forastieri:2017oma,Chu:2018gxk}. While most of the past literature has focused on heavy mediators,\footnote{Note that many of these papers assume entropy conservation during dark sector equilibriation, and as a result find $\Neff < \Neff^{SM}$. As this process is happening out-of-equilibrium, this assumption does not hold. In the case of number conserving interactions, the neutrinos develop a chemical potential such that $\Neff$ remains fixed.} we focus on the case of light mediators $m_\phi \ll \mdark$.

\enlargethispage*{3\baselineskip}
In our model we find that $\alpha_d=0$ (no coupling) is not the only problematic region for these cases.
Below we show that if the interaction is too strong ($\alpha_d \to 1$) the onset of SM neutrino free steaming is delayed and typically alters the CMB spectra considerably. We find that one is pushed to a fine-tuned corner of parameter space in order to accommodate the observations: the interaction strength should be strong enough to postpone thermalization until after BBN, while should be weak enough for the SM neutrinos to free stream for ample time before the production of the CMB. For concreteness, we start our exploration by focusing on the best fit values from MiniBoonE/MicroBoonE~\cite{MiniBooNE:2022emn}, $\mdark = 0.5\mathrm{eV}$ and a mixing angle of $\theta_0=0.09$.  We refer to this particular setting as \enquote{Anomaly}.

\pagebreak[20]

\begin{table}[H]
    \centering
    \begin{tabular}{c c|c c}
        Model & Dataset & Best-fit $\chi^2$ & Difference to $\Lambda$CDM\\ \hline \rule{0pt}{1em}
        \multirow{2}{*}{$\Lambda$CDM} & baseline & 4192.1 & $0$\\
        & baseline+SH0ES & 4102.3 & $0$\\ \hline \rule{0pt}{1em}
        \multirow{2}{*}{\wzdrnu ($\alphad=1$, narrow prior)} & baseline & 4191.5 & $-0.6$\\
        & baseline+SH0ES & 4102.0 & $-0.3$ \\ \hline \rule{0pt}{1em}
        \wzdrnu ($\alphad=1$, $1\nu$ interaction) & baseline & 4189.3 & $-1.8$\\ \hline \rule{0pt}{1em}
        \wzdrnu ($\alphad=1$, broad prior) & baseline & 4191.2 &  -0.9 \\ 
        \wzdrnu ($\alphad=1$, IDR limit) & baseline & 4195.2 & $+3.1$ \\
        \wzdrnu ($\alphad=1$, multimodal) & baseline & 4193.1 & $+1.0$ \\ \hline \rule{0pt}{1em}
        \multirow{2}{*}{Anomaly} & baseline & 4191.8 & $-0.3$\\
        & baseline+SH0ES & 4099.7 & $-2.6$\\ \hline \rule{0pt}{1em}
        \multirow{2}{*}
        {\wzdrnu (narrow prior)} & baseline & 4190.5 & $-1.6$\\
        & baseline+SH0ES & 4091.4& $-10.9$ \\ \hline \rule{0pt}{1em}
        \multirow{2}{*}{\wzdrnu} & baseline & 4190.3 & $-1.8$\\
        & baseline+SH0ES & 4087.6 & $-14.7$ \\ \hline \rule{0pt}{1em}
        \multirow{2}{*}{\wzdrnu ($1\nu$ interaction)} & baseline & 4191.3 & $-0.8$\\
        & baseline+SH0ES & 4093.4 & $-8.9$\\ \hline
    \end{tabular}
    \caption{Bestfit $\chi^2$ for different investigated models and comparison to $\Lambda$CDM. The anomaly case is described in \cref{ssec:weakly_coupled}. The minimized chisquares are trustable to around $\sigma (\chi^2) \simeq 0.3$. None of the models show a large improvement in the baseline case, while there are slight improvements in easing the Hubble tension when $\alphad$ is left to vary (see \cref{ssec:weakly_coupled}). The corresponding best fitting parameters are reported in \cref{tab:bestfits}.}
    \label{tab:chisquares}
\end{table}
\begin{figure}[H]
    \centering
    \includegraphics[width=0.68\textwidth]{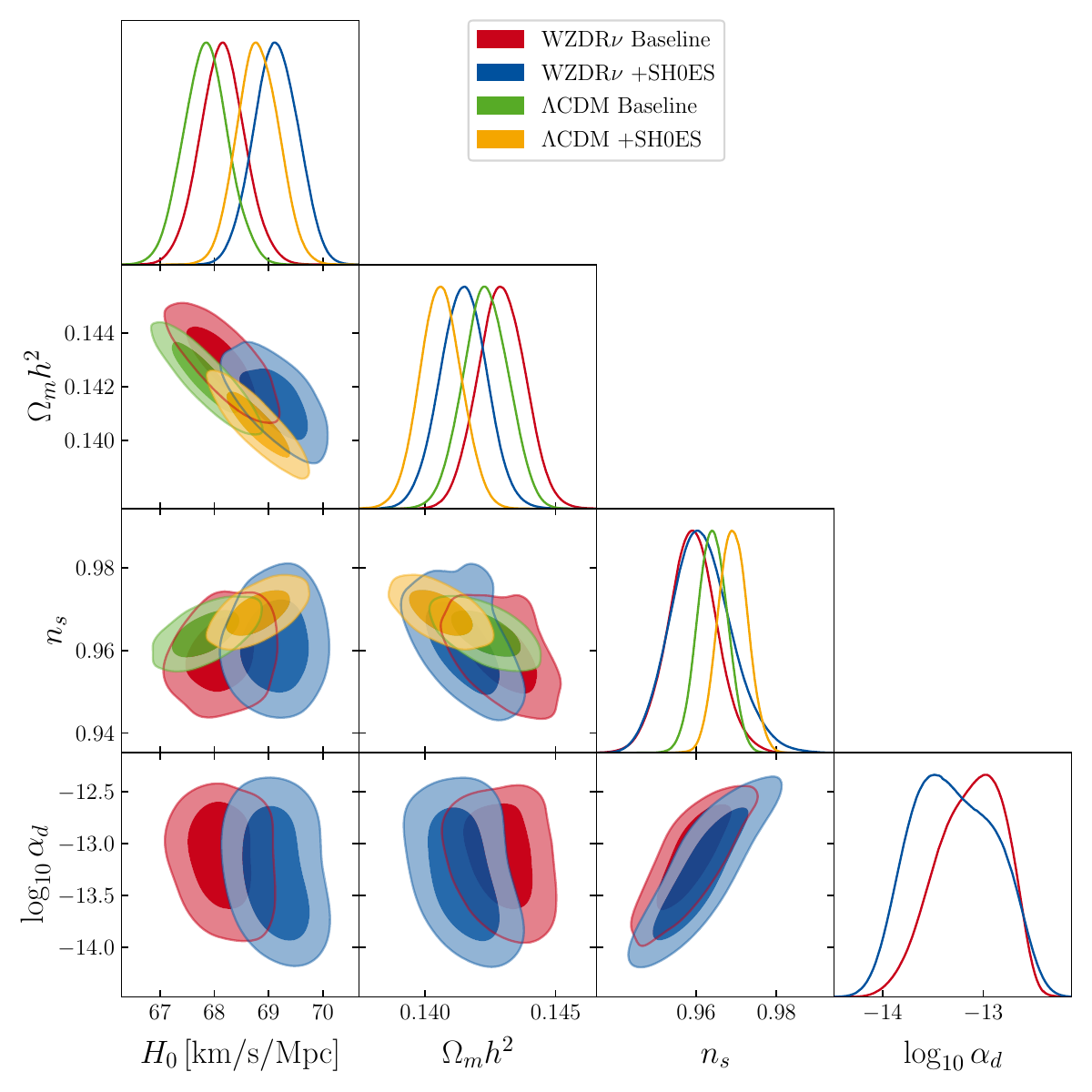}
    \caption{\label{fig:alpha_anomaly} One-dimensional posteriors and two-dimensional constraints at 68\% and 95\% CL for a range of parameters. Shown are the $\Lambda$CDM model and the \enquote{Anomaly} model (see text, $\theta_0 = 0.089$ and $\mdark = 0.54 \mathrm{eV}$) using baseline and baseline+SH0ES datasets. There is a BBN cutoff temperature at 100keV and a flat prior on $\log_{10}(\alphad) \in [-20,0]$.}
\end{figure}

\clearpage

We first investigate a model with the specific parameters of \cite{MiniBooNE:2022emn} and check the corresponding preferred parameters (like $\alphad$). We show the results in \cref{fig:alpha_anomaly}, where we can see that this part of the parameter space prefers small non-zero values of $\alpha_d$ around $10^{-14}$ to $10^{-13}$, as expected. 
\enlargethispage*{2\baselineskip}According to \cref{fig:alpha_dependence} these values correspond to small incomplete thermalization in the early universe and then also incomplete re-thermalization at lower redshifts. To be very explicit, we find that a model with the neutrino anomaly parameters of \cite{MiniBooNE:2022emn} can be compatible with CMB anisotropy observations and even be mildly helpful in terms of the Hubble tension. Comparing with \cref{tab:chisquares}, we find a very mild preference for such a model. At the same time this model is significantly more permissive in $n_s$ (allowing lower values, such as those compatible with \cite{Rogers+2023}), and could lead to differences in the power spectrum that might be observable with future galaxy or Lyman-$\alpha$ forest surveys, though a precise determination is left for future work.

\begin{figure}
    \centering
    \includegraphics[width=0.9\textwidth]{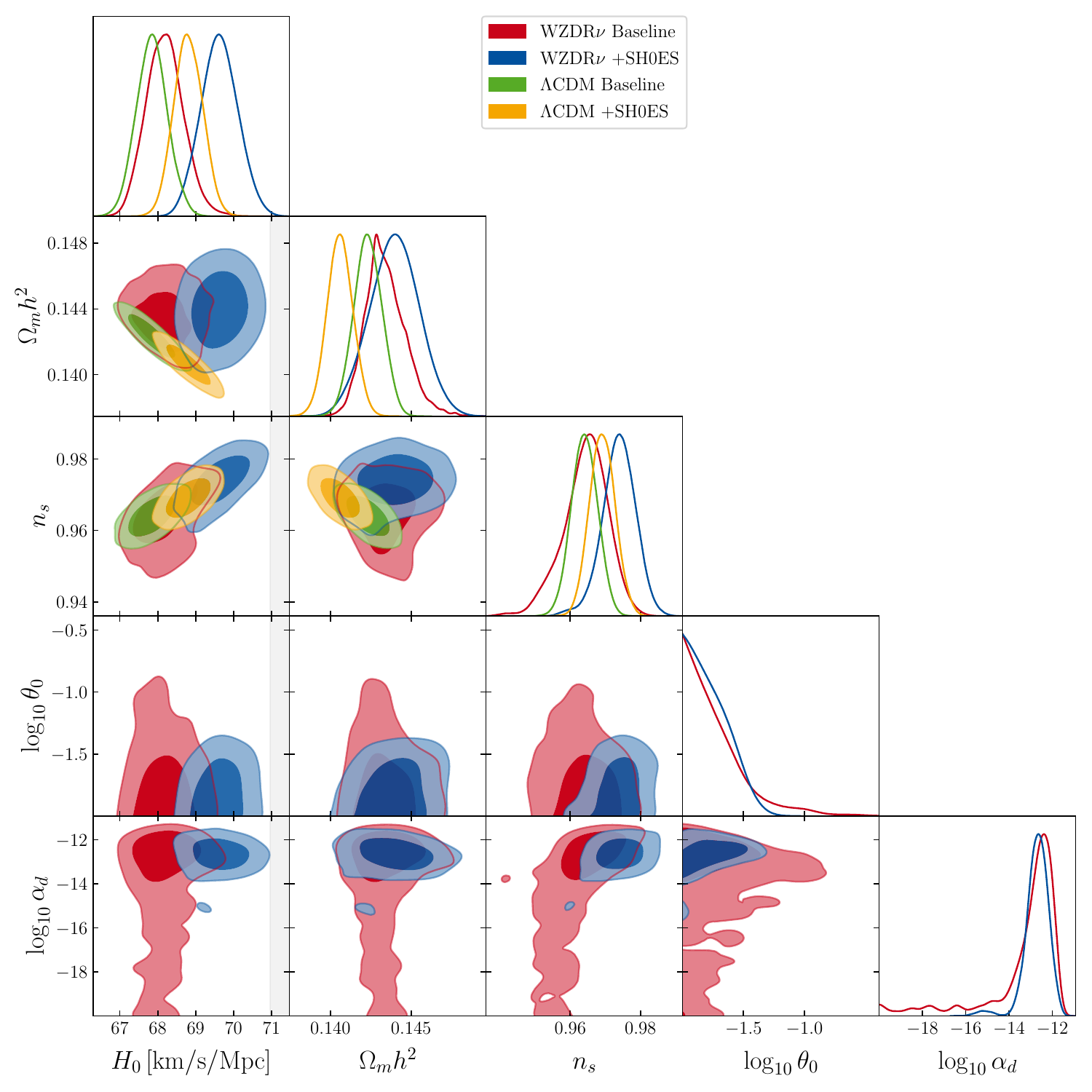}
    \caption{\label{fig:alpha_narrow} One-dimensional posteriors and two-dimensional constraints at 68\% and 95\% CL for a range of parameters. Shown are the $\Lambda$CDM model and the \wzdrnu model using the baseline and baseline+SH0ES datasets. In this case, priors on the \wzdrnu parameters are $\log_{10}\alphad\in[-20,0]$, $\log_{10}\theta_0\in[-2,0]$, and $\mdark\in[0.08,8]\,\mathrm{eV}$ (linear prior). The 95\% confidence interval for the SH0ES measurement of $H_0$ is shown by the light gray shaded region.
    }
\end{figure}

\begin{figure}
    \centering
    \includegraphics[width=0.9\textwidth]{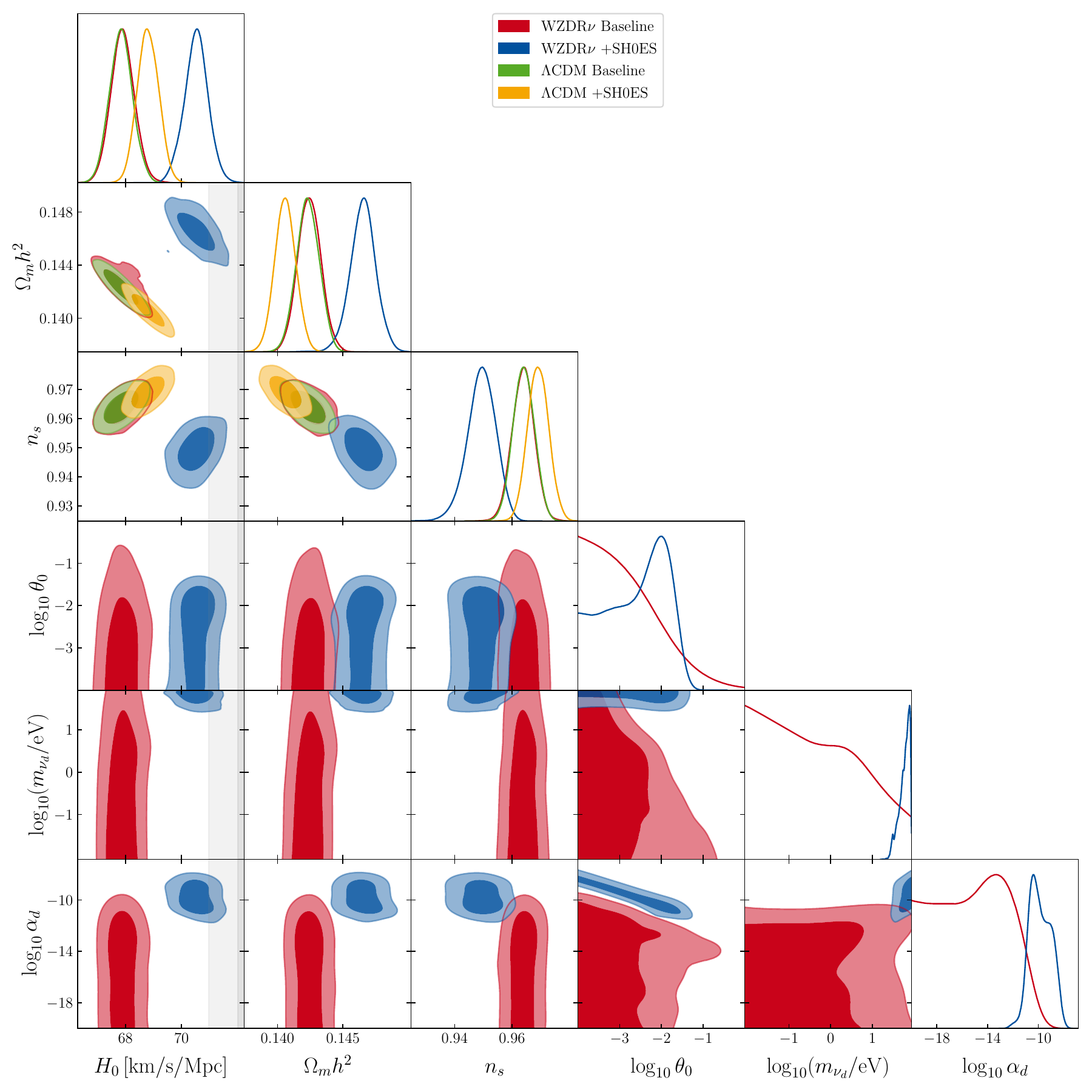}
    \caption{\label{fig:alpha_wide} One-dimensional posteriors and two-dimensional constraints at 68\% and 95\% CL for a range of parameters. Shown are the $\Lambda$CDM model and the \wzdrnu model using the baseline and baseline+SH0ES datasets. In this case, priors on the \wzdrnu parameters are $\log_{10}\alphad\in[-20,0]$, $\log_{10}\theta_0\in[-2,0]$, and $\log_{10}(\mdark/\mbox{eV})\in[-2.1,1.9]$. The 68\% and 95\% confidence intervals for the SH0ES measurement of $H_0$ are shown by the gray and lighter gray shaded regions.
    }
\end{figure}

\begin{figure}
    \centering
    \includegraphics[width=0.9\textwidth]{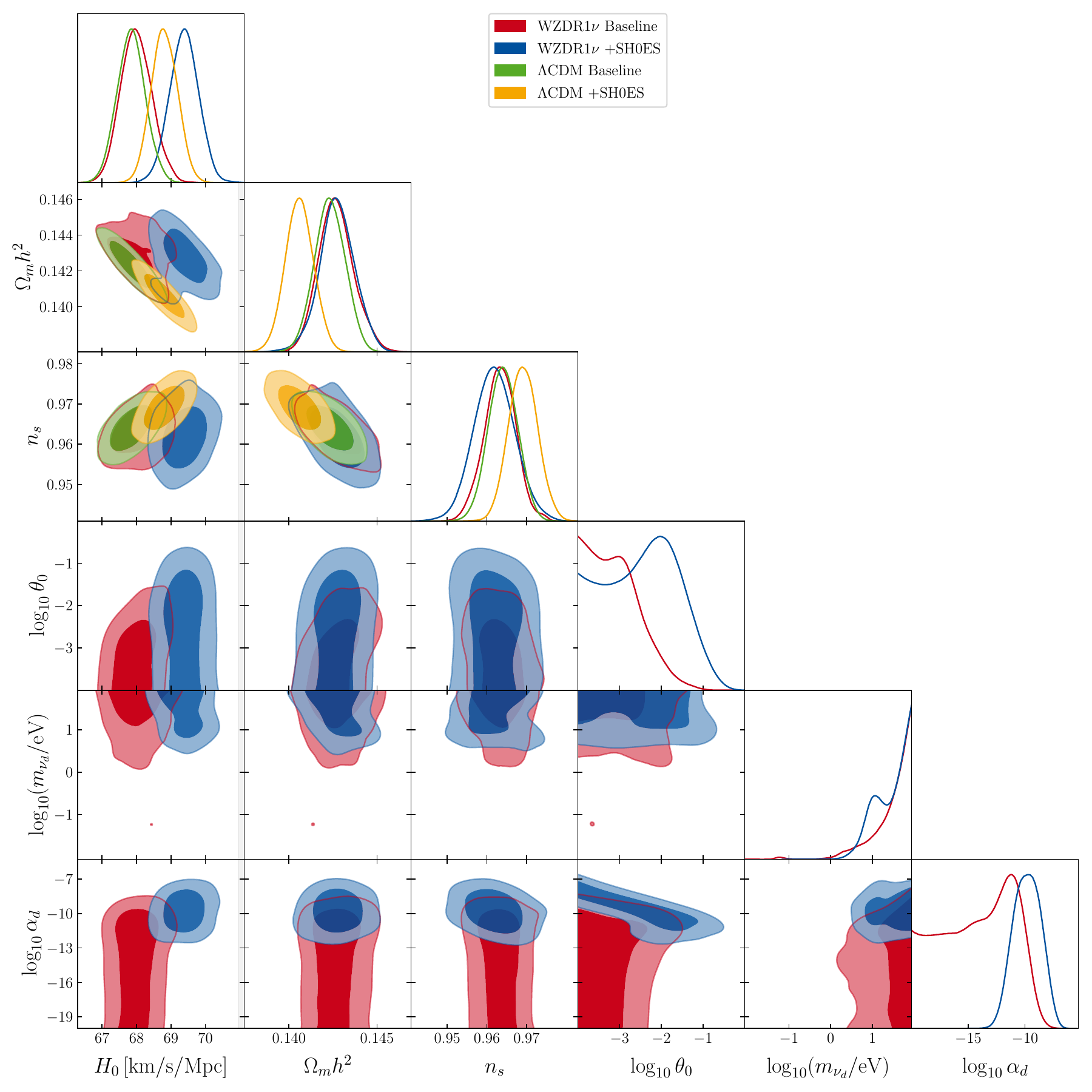}
    \caption{\label{fig:alpha_wide_1nu}  One-dimensional posteriors and two-dimensional constraints at 68\% and 95\% CL for a range of parameters. Shown are the $\Lambda$CDM model and the \wzdrnu mode, with only one standard model neutrino interacting, using the baseline and baseline+SH0ES datasets. In this case, priors on the \wzdrnu parameters are $\log_{10}\alphad\in[-20,0]$, $\log_{10}\theta_0\in[-2,0]$, and $\log_{10}(\mdark/\mbox{eV})\in[-2.1,1.9]$ (note that the prior on $\mdark$ is implemented linearly in this case).}
\end{figure}

Inspired by the decent performance of this specific model, we extend the parameter space to include a larger parameter range of the mixing angle $\theta_0$ and fermion mass $\mdark$ in order to probe also other parameter combinations that have been shown to be viable candidates for the neutrino anomalies \cite{Kopp+2013}. We extend the priors to $\log_{10}(\theta_0) \in [-2,0]$ and a linear prior on $\mdark \in [0.08,8]\mathrm{eV}$. This case is shown in \cref{fig:alpha_narrow} (where we once again marginalize over and do not show the mass explicitly as it does not display any interesting correlations). In this case we find that there is a broad allowance on $\alphad$ if no $H_0$ prior is imposed. However, there still is a slight preference for values around $\alpha_d \sim 10^{-13}$ from the CMB data, since this region of large $\theta_0>10^{-2}$ requires at least a bit suppression to avoid complete Dodelson-Widrow thermalization. In this case the posterior is $H_0 = 69.63 \pm0.49$ km/s/Mpc with SH0ES, leading to a Gaussian tension of $3.0\sigma$. If a prior on the Hubble parameter is imposed, this selects a region of the parameter space where there is an incomplete thermalization before/during BBN and the model mostly behaves like a decoupled WZDR model otherwise. \cref{tab:chisquares} shows that this slightly broader parameter space allows for much better solutions to the Hubble tension, being preferred by a $\Delta \chi^2 \simeq -11$ compared to $\Lambda$CDM when the $H_0$ prior is imposed.

\enlargethispage*{1\baselineskip}
Motivated by the preference for smaller $\theta_0$ values towards the edge of the prior that we observe in \cref{fig:alpha_narrow}, we also extend the priors even more to $\log_{10}(\theta_0) \in [-4,0]$ and $\log_{10}(\mdark/\mathrm{eV}) \in [-2.1,1.9]$. The corresponding results are shown in \cref{fig:alpha_wide}. In this extended parameter space, we find full compatibility of smaller $\alpha_d \lesssim 10^{-12}$ for correspondingly smaller values of $\theta_0 \lesssim 10^{-2}$ that would not lead to full Dodelson-Widrow thermalization even for $\alpha_d \to 0$. The main constraint on $\alpha_d$ seems to stem in this case from a consideration of not strongly coupling at a too early time (in order not to be strongly coupled during or before recombination).

Once a prior on the Hubble parameter is imposed, a slightly different region in parameter space is preferred. 
A somewhat larger value of $\theta_0$ coupled with a large value of $\alpha_d$ is preferred, introducing an incomplete thermalization that generates a small initial abundance. This small initial abundance that is generated without fully coupling then undergoes the usual WZDR transition at a later stage shortly before recombination, helping ease the Hubble tension (and thus being preferred once the SH0ES prior is added). Regions with small $\alpha$ but $\theta_0$ just right to only partially thermalize at BBN are also slightly preferred for the same reason. Notably, this case also performs quite well in terms of easing the Hubble tension as can be gleaned from \cref{tab:chisquares}, which shows a significant improvement of the bestfit $\chi^2$ compared to $\Lambda$CDM when the Hubble prior is imposed. With 
$70.53 \pm 0.41 \mathrm{km/s/Mpc}$ in this case, the Gaussian tension is reduced to around $2.2\sigma$.

\enlargethispage*{1\baselineskip}
It should be noted that we do not restrict the model at BBN beyond enforcing $\Gamma/H < 1$ at the end of BBN. As such, these results are conservative in constraining the model and tighter constraints could possibly be derived from BBN. For recent analysis of BBN constraints in fully thermalizing models we refer the reader to~\cite{Giovanetti:2024orj}, while a more detailed analysis for partial thermalization is left for future work.

We also investigate a case in which only one of the neutrinos is interacting, but this regime has a weaker cosmological impact (due to fewer neutrinos interacting), leading to $H_0=69.38 \pm 0.42 \mathrm{km/s/Mpc}$ ($3.3\sigma$ tension) and a much smaller difference in $\chi^2_\mathrm{eff} = - 2\ln L$. We show these constraints for completeness in \cref{fig:alpha_wide_1nu}.

\section{Conclusions}\label{sec:conclusions}

\enlargethispage*{2\baselineskip}
We have presented a new model of neutrinos mixing with a dark sector that we call \wzdrnu. This model has both a well motivated description of the initial thermalization with plenty of parameter space that can avoid thermalization during or before big bang nucleosynthesis (BBN), as well as a mass step before recombination that allows for an enhancement of the dark radiation abundance and corresponding reduction of the sound horizon in view of easing the Hubble tension. For this model, based on the initial idea presented in \cite{Aloni:2023tff}, we have derived the full background and perturbation evolution equations and implemented them within the \texttt{CLASS} code.

The model has two very relevant limits. In the strongly coupled limit the self-interaction strength ($\alphad$) is large (of order unity). We first investigate intermediate-mass ($m_\nudark \sim \mathcal{O}(\mathrm{eV})$) models, which are motivated by their mass-step happening at a time that would inject additional dark radiation shortly before recombination (steppped dark radiation), which has proven in \cite{Aloni:2021eaq} to be efficient in easing the Hubble tension.  Surprisingly, in this regime we find that the model is strongly constrained to be close to the $\Lambda$CDM limit of small interaction strength due to the additional perturbative effects of the model. As such, despite at the background level being very similar to aforementioned stepped dark radiation models, the perturbative effects prevent the model from being successful in this regime. This remains true for single-neutrino interactions. However, a broader parameter space is opened up once a wider range of masses is allowed. In particular the low and high mass cases allow larger couplings. While in the low mass region the coupling always has to be small enough to thermalize essentially after recombination, in the high mass region thermalization before recombination is possible with only a slight penalty in the overall fit. This somewhat disconnected region of parameter space is essentially equivalent to a combination of strongly self interacting and free streaming dark radiation, similar to what has been investigated in \cite{Blinov+2020}. This part of the parameter space is also somewhat more efficient in easing the Hubble tension (with $H_0 = 69.64 \pm 0.44 \mathrm{km/s/Mpc}$, a $3\sigma$ Gaussian tension), albeit being less preferred by a difference in effective chi-square of $\Delta \chi^2 = +3.1$ with respect to $\Lambda$CDM (whereas the low-mass region achieves $\Delta \chi^2 = -0.9$, both without an $H_0$ prior).

The other limit is the weakly coupled limit, for which a smaller self-interaction strength $\alphad$ can lead to interesting dynamics of only partially thermalizing both in the early (pre-BBN) and in the late universe. Additionally, this class of models can accomodate larger mixing angles, which are interesting to explore with respect to anomalies in neutrino oscillations (such as measured by \cite{MiniBooNE:2022emn}, see \cite{Dasgupta:2021ies} for a review). We find that the \wzdrnu model in its weakly coupled limit can accommodate such anomalous neutrino mixing angles while remaining cosmologically undetectable, or, even more interestingly, even aiding mildly in easing the Hubble tension.

The partial thermalization in the early and late universe also unlocks new parts of the parameter space, however, which are even more interesting in this regard. We find that if larger masses are probed, the partial thermalization regime can have a strong impact on the Hubble parameter (giving $H_0=70.53\pm 0.41 \mathrm{km/s/Mpc}$, a $2.2\sigma$ tension with the SH0ES measurement) while remaining excellently compatible with the CMB anisotropy observations (resulting in a total $\Delta \chi^2 = -14.7$ compared to $\Lambda$CDM, which would be mildly favored in terms of the Aikaike information criterion, or other model comparison assessments). We stress here that a caveat to these conclusions is that we do not fully model the impact of a partially thermalizing \wzdrnu model on BBN observables, which is left for future work.

Similarly, there are multiple avenues to explore with regard to more fully investigating this interesting model:
\begin{itemize}
    \itemsep0em
    \item As we observe in the strong coupling regime that the model has at least two local minima that provide good fits to the CMB, further investigation using nested sampling might reveal new interesting regimes of parameter space, both for the strong and the weak coupling regimes.
    \item Deriving a more complete description of non-relativistic neutrino oscillations, which would allow for a description of the interaction strength beyond a phenomelogical factor. This would also allow for the code to follow the full momentum-dependent phase-space distributions of the individual species (treating all neutrinos as massive). Both of these changes are not expected to have a large impact on the final result, but this should be confirmed explicitly.
    \item Investigating the impact of the model on different observables characterized by additional datasets, such as for example ACT/SPT \cite{ACT:2020gnv,SPT-3G:2021vps} CMB anisotropy data as well as the new BAO results from DESI \cite{DESI:2016fyo,DESI:2024lzq,DESI:2024mwx,DESI:2024uvr}.
    \item Performing a more complete treatment of the impact of the partially thermalizing parameter space of the model on BBN, as mentioned above.
\end{itemize}

The results in this work, we believe, encourage future study of the scenario we have outlined and other scenarios of dark radiation models that may have impacts on the Hubble tension and our understanding of cosmology.

\section*{Acknowledgements}
We thank Yuval Grossman, Martin Schmaltz, Yael Shadmi, and Neal Weiner for useful discussions and comments. We acknowledge use of the Tufts HPC research cluster. IJA is supported by NASA grant 80NSSC22K081 and partly supported by the John
F. Burlingame Graduate Fellowship in Physics at Tufts University.
The work of DA is supported by the U.S. Department of Energy (DOE) under Award DE-SC0015845.
NS acknowledges the support of the following Maria de Maetzu fellowship grant: Esta publicaci\'on es parte de la ayuda \mbox{CEX2019-000918-M}, financiada por MCIN/AEI/10.13039/501100011033. NS also acknowledges support by MICINN grant number PID \mbox{2022-141125NB-I00}.

\bibliographystyle{utphys}
\bibliography{stepped_interactin_neutrino}

\providecommand{\href}[2]{#2}\begingroup\raggedright\begin{thebibliography}{10}

\bibitem{Riess:2021jrx}
A.~G. Riess {\em et al.}, ``{A Comprehensive Measurement of the Local Value of
  the Hubble Constant with 1 km s$^{-1}$ Mpc$^{-1}$ Uncertainty from the Hubble
  Space Telescope and the SH0ES Team},''
  \href{http://dx.doi.org/10.3847/2041-8213/ac5c5b}{{\em Astrophys. J. Lett.}
  {\bf 934} (2022) no.~1, L7}, \href{http://arxiv.org/abs/2112.04510}{{\tt
  arXiv:2112.04510 [astro-ph.CO]}}.

\bibitem{Planck:2018vyg}
{\bf Planck} Collaboration, N.~Aghanim {\em et al.}, ``{Planck 2018 results.
  VI. Cosmological parameters},''
  \href{http://dx.doi.org/10.1051/0004-6361/201833910}{{\em Astron. Astrophys.}
  {\bf 641} (2020)  A6}, \href{http://arxiv.org/abs/1807.06209}{{\tt
  arXiv:1807.06209 [astro-ph.CO]}}. [Erratum: Astron.Astrophys. 652, C4
  (2021)].

\bibitem{Schoneberg:2021qvd}
N.~Sch\"oneberg, G.~Franco~Abell\'an, A.~P\'erez~S\'anchez, S.~J. Witte,
  V.~Poulin, and J.~Lesgourgues, ``{The H0 Olympics: A fair ranking of proposed
  models},'' \href{http://dx.doi.org/10.1016/j.physrep.2022.07.001}{{\em Phys.
  Rept.} {\bf 984} (2022)  1--55}, \href{http://arxiv.org/abs/2107.10291}{{\tt
  arXiv:2107.10291 [astro-ph.CO]}}.

\bibitem{DiValentino:2021izs}
E.~Di~Valentino, O.~Mena, S.~Pan, L.~Visinelli, W.~Yang, A.~Melchiorri, D.~F.
  Mota, A.~G. Riess, and J.~Silk, ``{In the realm of the Hubble
  tension\textemdash{}a review of solutions},''
  \href{http://dx.doi.org/10.1088/1361-6382/ac086d}{{\em Class. Quant. Grav.}
  {\bf 38} (2021) no.~15, 153001}, \href{http://arxiv.org/abs/2103.01183}{{\tt
  arXiv:2103.01183 [astro-ph.CO]}}.

\bibitem{Verde:2023lmm}
L.~Verde, N.~Sch\"oneberg, and H.~Gil-Mar\'\i{}n, ``{A tale of many $H_0$},''
  \href{http://arxiv.org/abs/2311.13305}{{\tt arXiv:2311.13305 [astro-ph.CO]}}.

\bibitem{Freedman:2021ahq}
W.~L. Freedman, ``{Measurements of the Hubble Constant: Tensions in
  Perspective},'' \href{http://dx.doi.org/10.3847/1538-4357/ac0e95}{{\em
  Astrophys. J.} {\bf 919} (2021) no.~1, 16},
  \href{http://arxiv.org/abs/2106.15656}{{\tt arXiv:2106.15656 [astro-ph.CO]}}.

\bibitem{Freedman:2023jcz}
W.~L. Freedman and B.~F. Madore, ``{Progress in direct measurements of the
  Hubble constant},''
  \href{http://dx.doi.org/10.1088/1475-7516/2023/11/050}{{\em JCAP} {\bf 11}
  (2023)  050}, \href{http://arxiv.org/abs/2309.05618}{{\tt arXiv:2309.05618
  [astro-ph.CO]}}.

\bibitem{Kamionkowski:2022pkx}
M.~Kamionkowski and A.~G. Riess, ``{The Hubble Tension and Early Dark
  Energy},'' \href{http://dx.doi.org/10.1146/annurev-nucl-111422-024107}{{\em
  Ann. Rev. Nucl. Part. Sci.} {\bf 73} (2023)  153--180},
  \href{http://arxiv.org/abs/2211.04492}{{\tt arXiv:2211.04492 [astro-ph.CO]}}.

\bibitem{Chacko:2015noa}
Z.~Chacko, Y.~Cui, S.~Hong, and T.~Okui, ``{Hidden dark matter sector, dark
  radiation, and the CMB},''
  \href{http://dx.doi.org/10.1103/PhysRevD.92.055033}{{\em Phys. Rev. D} {\bf
  92} (2015)  055033}, \href{http://arxiv.org/abs/1505.04192}{{\tt
  arXiv:1505.04192 [hep-ph]}}.

\bibitem{Buen-Abad:2015ova}
M.~A. Buen-Abad, G.~Marques-Tavares, and M.~Schmaltz, ``{Non-Abelian dark
  matter and dark radiation},''
  \href{http://dx.doi.org/10.1103/PhysRevD.92.023531}{{\em Phys. Rev. D} {\bf
  92} (2015) no.~2, 023531}, \href{http://arxiv.org/abs/1505.03542}{{\tt
  arXiv:1505.03542 [hep-ph]}}.

\bibitem{Chacko:2016kgg}
Z.~Chacko, Y.~Cui, S.~Hong, T.~Okui, and Y.~Tsai, ``{Partially Acoustic Dark
  Matter, Interacting Dark Radiation, and Large Scale Structure},''
  \href{http://dx.doi.org/10.1007/JHEP12(2016)108}{{\em JHEP} {\bf 12} (2016)
  108}, \href{http://arxiv.org/abs/1609.03569}{{\tt arXiv:1609.03569
  [astro-ph.CO]}}.

\bibitem{Cyr-Racine:2015ihg}
F.-Y. Cyr-Racine, K.~Sigurdson, J.~Zavala, T.~Bringmann, M.~Vogelsberger, and
  C.~Pfrommer, ``{ETHOS\textemdash{}an effective theory of structure formation:
  From dark particle physics to the matter distribution of the Universe},''
  \href{http://dx.doi.org/10.1103/PhysRevD.93.123527}{{\em Phys. Rev. D} {\bf
  93} (2016) no.~12, 123527}, \href{http://arxiv.org/abs/1512.05344}{{\tt
  arXiv:1512.05344 [astro-ph.CO]}}.

\bibitem{Lesgourgues:2015wza}
J.~Lesgourgues, G.~Marques-Tavares, and M.~Schmaltz, ``{Evidence for dark
  matter interactions in cosmological precision data?},''
  \href{http://dx.doi.org/10.1088/1475-7516/2016/02/037}{{\em JCAP} {\bf 02}
  (2016)  037}, \href{http://arxiv.org/abs/1507.04351}{{\tt arXiv:1507.04351
  [astro-ph.CO]}}.

\bibitem{Brust:2017nmv}
C.~Brust, Y.~Cui, and K.~Sigurdson, ``{Cosmological Constraints on Interacting
  Light Particles},''
  \href{http://dx.doi.org/10.1088/1475-7516/2017/08/020}{{\em JCAP} {\bf 08}
  (2017)  020}, \href{http://arxiv.org/abs/1703.10732}{{\tt arXiv:1703.10732
  [astro-ph.CO]}}.

\bibitem{Buen-Abad:2017gxg}
M.~A. Buen-Abad, M.~Schmaltz, J.~Lesgourgues, and T.~Brinckmann, ``{Interacting
  Dark Sector and Precision Cosmology},''
  \href{http://dx.doi.org/10.1088/1475-7516/2018/01/008}{{\em JCAP} {\bf 01}
  (2018)  008}, \href{http://arxiv.org/abs/1708.09406}{{\tt arXiv:1708.09406
  [astro-ph.CO]}}.

\bibitem{Archidiacono:2019wdp}
M.~Archidiacono, D.~C. Hooper, R.~Murgia, S.~Bohr, J.~Lesgourgues, and M.~Viel,
  ``{Constraining Dark Matter-Dark Radiation interactions with CMB, BAO, and
  Lyman-$\alpha$},''
  \href{http://dx.doi.org/10.1088/1475-7516/2019/10/055}{{\em JCAP} {\bf 10}
  (2019)  055}, \href{http://arxiv.org/abs/1907.01496}{{\tt arXiv:1907.01496
  [astro-ph.CO]}}.

\bibitem{Blinov:2020hmc}
N.~Blinov and G.~Marques-Tavares, ``{Interacting radiation after Planck and its
  implications for the Hubble Tension},''
  \href{http://dx.doi.org/10.1088/1475-7516/2020/09/029}{{\em JCAP} {\bf 09}
  (2020)  029}, \href{http://arxiv.org/abs/2003.08387}{{\tt arXiv:2003.08387
  [astro-ph.CO]}}.

\bibitem{Allali:2024cji}
I.~J. Allali, A.~Notari, and F.~Rompineve, ``{Dark Radiation with Baryon
  Acoustic Oscillations from DESI 2024 and the $H_0$ tension},''
  \href{http://arxiv.org/abs/2404.15220}{{\tt arXiv:2404.15220 [astro-ph.CO]}}.

\bibitem{Aloni:2021eaq}
D.~Aloni, A.~Berlin, M.~Joseph, M.~Schmaltz, and N.~Weiner, ``{A Step in
  understanding the Hubble tension},''
  \href{http://dx.doi.org/10.1103/PhysRevD.105.123516}{{\em Phys. Rev. D} {\bf
  105} (2022) no.~12, 123516}, \href{http://arxiv.org/abs/2111.00014}{{\tt
  arXiv:2111.00014 [astro-ph.CO]}}.

\bibitem{Joseph:2022jsf}
M.~Joseph, D.~Aloni, M.~Schmaltz, E.~N. Sivarajan, and N.~Weiner, ``{A Step in
  understanding the S8 tension},''
  \href{http://dx.doi.org/10.1103/PhysRevD.108.023520}{{\em Phys. Rev. D} {\bf
  108} (2023) no.~2, 023520}, \href{http://arxiv.org/abs/2207.03500}{{\tt
  arXiv:2207.03500 [astro-ph.CO]}}.

\bibitem{Schoneberg:2022grr}
N.~Sch\"oneberg and G.~Franco~Abell\'an, ``{A step in the right direction?
  Analyzing the Wess Zumino Dark Radiation solution to the Hubble tension},''
  \href{http://dx.doi.org/10.1088/1475-7516/2022/12/001}{{\em JCAP} {\bf 12}
  (2022)  001}, \href{http://arxiv.org/abs/2206.11276}{{\tt arXiv:2206.11276
  [astro-ph.CO]}}.

\bibitem{Allali:2023zbi}
I.~J. Allali, F.~Rompineve, and M.~P. Hertzberg, ``{Dark sectors with mass
  thresholds face cosmological datasets},''
  \href{http://dx.doi.org/10.1103/PhysRevD.108.023527}{{\em Phys. Rev. D} {\bf
  108} (2023) no.~2, 023527}, \href{http://arxiv.org/abs/2305.14166}{{\tt
  arXiv:2305.14166 [astro-ph.CO]}}.

\bibitem{Aloni:2023tff}
D.~Aloni, M.~Joseph, M.~Schmaltz, and N.~Weiner, ``{Dark Radiation from
  Neutrino Mixing after Big Bang Nucleosynthesis},''
  \href{http://arxiv.org/abs/2301.10792}{{\tt arXiv:2301.10792 [astro-ph.CO]}}.

\bibitem{Dolgov:1980cq}
A.~D. Dolgov, ``{Neutrinos in the Early Universe},'' {\em Sov. J. Nucl. Phys.}
  {\bf 33} (1981)  700--706.

\bibitem{Barbieri:1989ti}
R.~Barbieri and A.~Dolgov, ``{Bounds on Sterile-neutrinos from
  Nucleosynthesis},''
  \href{http://dx.doi.org/10.1016/0370-2693(90)91203-N}{{\em Phys. Lett. B}
  {\bf 237} (1990)  440--445}.

\bibitem{Barbieri:1990vx}
R.~Barbieri and A.~Dolgov, ``{Neutrino oscillations in the early universe},''
  \href{http://dx.doi.org/10.1016/0550-3213(91)90396-F}{{\em Nucl. Phys. B}
  {\bf 349} (1991)  743--753}.

\bibitem{Enqvist:1990ad}
K.~Enqvist, K.~Kainulainen, and J.~Maalampi, ``{Refraction and Oscillations of
  Neutrinos in the Early Universe},''
  \href{http://dx.doi.org/10.1016/0550-3213(91)90397-G}{{\em Nucl. Phys. B}
  {\bf 349} (1991)  754--790}.

\bibitem{Sigl:1993ctk}
G.~Sigl and G.~Raffelt, ``{General kinetic description of relativistic mixed
  neutrinos},'' \href{http://dx.doi.org/10.1016/0550-3213(93)90175-O}{{\em
  Nucl. Phys. B} {\bf 406} (1993)  423--451}.

\bibitem{McKellar:1992ja}
B.~H.~J. McKellar and M.~J. Thomson, ``{Oscillating doublet neutrinos in the
  early universe},'' \href{http://dx.doi.org/10.1103/PhysRevD.49.2710}{{\em
  Phys. Rev. D} {\bf 49} (1994)  2710--2728}.

\bibitem{Dodelson:1993je}
S.~Dodelson and L.~M. Widrow, ``{Sterile-neutrinos as dark matter},''
  \href{http://dx.doi.org/10.1103/PhysRevLett.72.17}{{\em Phys. Rev. Lett.}
  {\bf 72} (1994)  17--20}, \href{http://arxiv.org/abs/hep-ph/9303287}{{\tt
  arXiv:hep-ph/9303287}}.

\bibitem{Schoeneberg2024}
N.~{Sch{\"o}neberg}, \href{http://dx.doi.org/10.48550/arXiv.2401.15054}{``{The
  2024 BBN baryon abundance update},''{\em arXiv e-prints} (Jan., 2024)
  arXiv:2401.15054}, \href{http://arxiv.org/abs/2401.15054}{{\tt
  arXiv:2401.15054 [astro-ph.CO]}}.

\bibitem{ParticleDataGroup:2022pth}
{\bf Particle Data Group} Collaboration, R.~L. Workman {\em et al.}, ``{Review
  of Particle Physics},'' \href{http://dx.doi.org/10.1093/ptep/ptac097}{{\em
  PTEP} {\bf 2022} (2022)  083C01}.

\bibitem{Shi:1998km}
X.-D. Shi and G.~M. Fuller, ``{A New dark matter candidate: Nonthermal sterile
  neutrinos},'' \href{http://dx.doi.org/10.1103/PhysRevLett.82.2832}{{\em Phys.
  Rev. Lett.} {\bf 82} (1999)  2832--2835},
  \href{http://arxiv.org/abs/astro-ph/9810076}{{\tt arXiv:astro-ph/9810076}}.

\bibitem{Ma:1995ey}
C.-P. Ma and E.~Bertschinger, ``{Cosmological perturbation theory in the
  synchronous and conformal Newtonian gauges},''
  \href{http://dx.doi.org/10.1086/176550}{{\em Astrophys. J.} {\bf 455} (1995)
  7--25}, \href{http://arxiv.org/abs/astro-ph/9506072}{{\tt
  arXiv:astro-ph/9506072}}.

\bibitem{Lesgourgues:2011re}
J.~Lesgourgues, ``{The Cosmic Linear Anisotropy Solving System (CLASS) I:
  Overview},'' \href{http://arxiv.org/abs/1104.2932}{{\tt arXiv:1104.2932
  [astro-ph.IM]}}.

\bibitem{Blas:2011rf}
D.~Blas, J.~Lesgourgues, and T.~Tram, ``{The Cosmic Linear Anisotropy Solving
  System (CLASS) II: Approximation schemes},''
  \href{http://dx.doi.org/10.1088/1475-7516/2011/07/034}{{\em JCAP} {\bf 07}
  (2011)  034}, \href{http://arxiv.org/abs/1104.2933}{{\tt arXiv:1104.2933
  [astro-ph.CO]}}.

\bibitem{Audren:2012wb}
B.~Audren, J.~Lesgourgues, K.~Benabed, and S.~Prunet, ``{Conservative
  Constraints on Early Cosmology: an illustration of the Monte Python
  cosmological parameter inference code},''
  \href{http://dx.doi.org/10.1088/1475-7516/2013/02/001}{{\em JCAP} {\bf 02}
  (2013)  001}, \href{http://arxiv.org/abs/1210.7183}{{\tt arXiv:1210.7183
  [astro-ph.CO]}}.

\bibitem{Brinckmann:2018cvx}
T.~Brinckmann and J.~Lesgourgues, ``{MontePython 3: boosted MCMC sampler and
  other features},'' \href{http://dx.doi.org/10.1016/j.dark.2018.100260}{{\em
  Phys. Dark Univ.} {\bf 24} (2019)  100260},
  \href{http://arxiv.org/abs/1804.07261}{{\tt arXiv:1804.07261 [astro-ph.CO]}}.

\bibitem{Aghanim:2019ame}
{\bf Planck} Collaboration, N.~Aghanim {\em et al.}, ``{Planck 2018 results. V.
  CMB power spectra and likelihoods},''
  \href{http://dx.doi.org/10.1051/0004-6361/201936386}{{\em Astron. Astrophys.}
  {\bf 641} (2020)  A5}, \href{http://arxiv.org/abs/1907.12875}{{\tt
  arXiv:1907.12875 [astro-ph.CO]}}.

\bibitem{Beutler:2011hx}
F.~Beutler, C.~Blake, M.~Colless, D.~H. Jones, L.~Staveley-Smith, L.~Campbell,
  Q.~Parker, W.~Saunders, and F.~Watson, ``{The 6dF Galaxy Survey: Baryon
  Acoustic Oscillations and the Local Hubble Constant},''
  \href{http://dx.doi.org/10.1111/j.1365-2966.2011.19250.x}{{\em Mon. Not. Roy.
  Astron. Soc.} {\bf 416} (2011)  3017--3032},
  \href{http://arxiv.org/abs/1106.3366}{{\tt arXiv:1106.3366 [astro-ph.CO]}}.

\bibitem{Ross:2014qpa}
A.~J. Ross, L.~Samushia, C.~Howlett, W.~J. Percival, A.~Burden, and M.~Manera,
  ``{The clustering of the SDSS DR7 main Galaxy sample \textendash{} I. A 4 per
  cent distance measure at $z = 0.15$},''
  \href{http://dx.doi.org/10.1093/mnras/stv154}{{\em Mon. Not. Roy. Astron.
  Soc.} {\bf 449} (2015) no.~1, 835--847},
  \href{http://arxiv.org/abs/1409.3242}{{\tt arXiv:1409.3242 [astro-ph.CO]}}.

\bibitem{Alam:2016hwk}
{\bf BOSS} Collaboration, S.~Alam {\em et al.}, ``{The clustering of galaxies
  in the completed SDSS-III Baryon Oscillation Spectroscopic Survey:
  cosmological analysis of the DR12 galaxy sample},''
  \href{http://dx.doi.org/10.1093/mnras/stx721}{{\em Mon. Not. Roy. Astron.
  Soc.} {\bf 470} (2017) no.~3, 2617--2652},
  \href{http://arxiv.org/abs/1607.03155}{{\tt arXiv:1607.03155 [astro-ph.CO]}}.

\bibitem{Brout:2022vxf}
D.~Brout {\em et al.}, ``{The Pantheon+ Analysis: Cosmological Constraints},''
  \href{http://dx.doi.org/10.3847/1538-4357/ac8e04}{{\em Astrophys. J.} {\bf
  938} (2022) no.~2, 110}, \href{http://arxiv.org/abs/2202.04077}{{\tt
  arXiv:2202.04077 [astro-ph.CO]}}.

\bibitem{Lewis:2019xzd}
A.~Lewis, ``{GetDist: a Python package for analysing Monte Carlo samples},''
  \href{http://arxiv.org/abs/1910.13970}{{\tt arXiv:1910.13970 [astro-ph.IM]}}.
\url{https://getdist.readthedocs.io}.
%%CITATION = ARXIV:1910.13970;%%.

\bibitem{Blinov+2020}
N.~{Blinov} and G.~{Marques-Tavares},
  \href{http://dx.doi.org/10.1088/1475-7516/2020/09/029}{``{Interacting
  radiation after Planck and its implications for the Hubble tension},''{\em
  JCAP} {\bf 2020} (Sept., 2020)  029},
  \href{http://arxiv.org/abs/2003.08387}{{\tt arXiv:2003.08387 [astro-ph.CO]}}.

\bibitem{Cyr-Racine:2013jua}
F.-Y. Cyr-Racine and K.~Sigurdson, ``{Limits on Neutrino-Neutrino Scattering in
  the Early Universe},''
  \href{http://dx.doi.org/10.1103/PhysRevD.90.123533}{{\em Phys. Rev. D} {\bf
  90} (2014) no.~12, 123533}, \href{http://arxiv.org/abs/1306.1536}{{\tt
  arXiv:1306.1536 [astro-ph.CO]}}.

\bibitem{Oldengott:2014qra}
I.~M. Oldengott, C.~Rampf, and Y.~Y.~Y. Wong, ``{Boltzmann hierarchy for
  interacting neutrinos I: formalism},''
  \href{http://dx.doi.org/10.1088/1475-7516/2015/04/016}{{\em JCAP} {\bf 04}
  (2015)  016}, \href{http://arxiv.org/abs/1409.1577}{{\tt arXiv:1409.1577
  [astro-ph.CO]}}.

\bibitem{Oldengott:2017fhy}
I.~M. Oldengott, T.~Tram, C.~Rampf, and Y.~Y.~Y. Wong, ``{Interacting neutrinos
  in cosmology: exact description and constraints},''
  \href{http://dx.doi.org/10.1088/1475-7516/2017/11/027}{{\em JCAP} {\bf 11}
  (2017)  027}, \href{http://arxiv.org/abs/1706.02123}{{\tt arXiv:1706.02123
  [astro-ph.CO]}}.

\bibitem{Kreisch:2019yzn}
C.~D. Kreisch, F.-Y. Cyr-Racine, and O.~Dor\'e, ``{Neutrino puzzle: Anomalies,
  interactions, and cosmological tensions},''
  \href{http://dx.doi.org/10.1103/PhysRevD.101.123505}{{\em Phys. Rev. D} {\bf
  101} (2020) no.~12, 123505}, \href{http://arxiv.org/abs/1902.00534}{{\tt
  arXiv:1902.00534 [astro-ph.CO]}}.

\bibitem{Das:2020xke}
A.~Das and S.~Ghosh, ``{Flavor-specific interaction favors strong neutrino
  self-coupling in the early universe},''
  \href{http://dx.doi.org/10.1088/1475-7516/2021/07/038}{{\em JCAP} {\bf 07}
  (2021)  038}, \href{http://arxiv.org/abs/2011.12315}{{\tt arXiv:2011.12315
  [astro-ph.CO]}}.

\bibitem{Escudero:2019gvw}
M.~Escudero and S.~J. Witte, ``{A CMB search for the neutrino mass mechanism
  and its relation to the Hubble tension},''
  \href{http://dx.doi.org/10.1140/epjc/s10052-020-7854-5}{{\em Eur. Phys. J. C}
  {\bf 80} (2020) no.~4, 294}, \href{http://arxiv.org/abs/1909.04044}{{\tt
  arXiv:1909.04044 [astro-ph.CO]}}.

\bibitem{EscuderoAbenza:2020egd}
M.~Escudero~Abenza and S.~J. Witte, ``{Could the Hubble Tension be Pointing
  Towards the Neutrino Mass Mechanism?},'' in {\em {Prospects in Neutrino
  Physics}}.
\newblock 4, 2020.
\newblock \href{http://arxiv.org/abs/2004.01470}{{\tt arXiv:2004.01470
  [hep-ph]}}.

\bibitem{Escudero:2021rfi}
M.~Escudero and S.~J. Witte, ``{The hubble tension as a hint of leptogenesis
  and neutrino mass generation},''
  \href{http://dx.doi.org/10.1140/epjc/s10052-021-09276-5}{{\em Eur. Phys. J.
  C} {\bf 81} (2021) no.~6, 515}, \href{http://arxiv.org/abs/2103.03249}{{\tt
  arXiv:2103.03249 [hep-ph]}}.

\bibitem{Sandner:2023ptm}
S.~Sandner, M.~Escudero, and S.~J. Witte, ``{Precision CMB constraints on
  eV-scale bosons coupled to neutrinos},''
  \href{http://dx.doi.org/10.1140/epjc/s10052-023-11864-6}{{\em Eur. Phys. J.
  C} {\bf 83} (2023) no.~8, 709}, \href{http://arxiv.org/abs/2305.01692}{{\tt
  arXiv:2305.01692 [hep-ph]}}.

\bibitem{Dasgupta:2021ies}
B.~Dasgupta and J.~Kopp, ``{Sterile Neutrinos},''
  \href{http://dx.doi.org/10.1016/j.physrep.2021.06.002}{{\em Phys. Rept.} {\bf
  928} (2021)  1--63}, \href{http://arxiv.org/abs/2106.05913}{{\tt
  arXiv:2106.05913 [hep-ph]}}.

\bibitem{Hamann:2011ge}
J.~Hamann, S.~Hannestad, G.~G. Raffelt, and Y.~Y.~Y. Wong, ``{Sterile neutrinos
  with eV masses in cosmology: How disfavoured exactly?},''
  \href{http://dx.doi.org/10.1088/1475-7516/2011/09/034}{{\em JCAP} {\bf 09}
  (2011)  034}, \href{http://arxiv.org/abs/1108.4136}{{\tt arXiv:1108.4136
  [astro-ph.CO]}}.

\bibitem{Hannestad:2013ana}
S.~Hannestad, R.~S. Hansen, and T.~Tram, ``{How Self-Interactions can Reconcile
  Sterile Neutrinos with Cosmology},''
  \href{http://dx.doi.org/10.1103/PhysRevLett.112.031802}{{\em Phys. Rev.
  Lett.} {\bf 112} (2014) no.~3, 031802},
  \href{http://arxiv.org/abs/1310.5926}{{\tt arXiv:1310.5926 [astro-ph.CO]}}.

\bibitem{Dasgupta:2013zpn}
B.~Dasgupta and J.~Kopp, ``{Cosmologically Safe eV-Scale Sterile Neutrinos and
  Improved Dark Matter Structure},''
  \href{http://dx.doi.org/10.1103/PhysRevLett.112.031803}{{\em Phys. Rev.
  Lett.} {\bf 112} (2014) no.~3, 031803},
  \href{http://arxiv.org/abs/1310.6337}{{\tt arXiv:1310.6337 [hep-ph]}}.

\bibitem{Mirizzi:2014ama}
A.~Mirizzi, G.~Mangano, O.~Pisanti, and N.~Saviano, ``{Collisional production
  of sterile neutrinos via secret interactions and cosmological
  implications},'' \href{http://dx.doi.org/10.1103/PhysRevD.91.025019}{{\em
  Phys. Rev. D} {\bf 91} (2015) no.~2, 025019},
  \href{http://arxiv.org/abs/1410.1385}{{\tt arXiv:1410.1385 [hep-ph]}}.

\bibitem{Tang:2014yla}
Y.~Tang, ``{More Is Different: Reconciling eV Sterile Neutrinos with
  Cosmological Mass Bounds},''
  \href{http://dx.doi.org/10.1016/j.physletb.2015.09.018}{{\em Phys. Lett. B}
  {\bf 750} (2015)  201--208}, \href{http://arxiv.org/abs/1501.00059}{{\tt
  arXiv:1501.00059 [hep-ph]}}.

\bibitem{Chu:2015ipa}
X.~Chu, B.~Dasgupta, and J.~Kopp, ``{Sterile neutrinos with secret
  interactions\textemdash{}lasting friendship with cosmology},''
  \href{http://dx.doi.org/10.1088/1475-7516/2015/10/011}{{\em JCAP} {\bf 10}
  (2015)  011}, \href{http://arxiv.org/abs/1505.02795}{{\tt arXiv:1505.02795
  [hep-ph]}}.

\bibitem{Forastieri:2017oma}
F.~Forastieri, M.~Lattanzi, G.~Mangano, A.~Mirizzi, P.~Natoli, and N.~Saviano,
  ``{Cosmic microwave background constraints on secret interactions among
  sterile neutrinos},''
  \href{http://dx.doi.org/10.1088/1475-7516/2017/07/038}{{\em JCAP} {\bf 07}
  (2017)  038}, \href{http://arxiv.org/abs/1704.00626}{{\tt arXiv:1704.00626
  [astro-ph.CO]}}.

\bibitem{Chu:2018gxk}
X.~Chu, B.~Dasgupta, M.~Dentler, J.~Kopp, and N.~Saviano, ``{Sterile neutrinos
  with secret interactions\textemdash{}cosmological discord?},''
  \href{http://dx.doi.org/10.1088/1475-7516/2018/11/049}{{\em JCAP} {\bf 11}
  (2018)  049}, \href{http://arxiv.org/abs/1806.10629}{{\tt arXiv:1806.10629
  [hep-ph]}}.

\bibitem{MiniBooNE:2022emn}
{\bf MiniBooNE} Collaboration, A.~A. Aguilar-Arevalo {\em et al.}, ``{MiniBooNE
  and MicroBooNE Combined Fit to a 3+1 Sterile Neutrino Scenario},''
  \href{http://dx.doi.org/10.1103/PhysRevLett.129.201801}{{\em Phys. Rev.
  Lett.} {\bf 129} (2022) no.~20, 201801},
  \href{http://arxiv.org/abs/2201.01724}{{\tt arXiv:2201.01724 [hep-ex]}}.

\bibitem{Rogers+2023}
K.~K. {Rogers} and V.~{Poulin},
  \href{http://dx.doi.org/10.48550/arXiv.2311.16377}{``{$5 \sigma$ tension
  between Planck cosmic microwave background and eBOSS Lyman-alpha forest and
  constraints on physics beyond $\Lambda$CDM},''{\em arXiv e-prints} (Nov.,
  2023)  arXiv:2311.16377}, \href{http://arxiv.org/abs/2311.16377}{{\tt
  arXiv:2311.16377 [astro-ph.CO]}}.

\bibitem{Kopp+2013}
J.~{Kopp}, P.~A.~N. {Machado}, M.~{Maltoni}, and T.~{Schwetz},
  \href{http://dx.doi.org/10.1007/JHEP05(2013)050}{``{Sterile neutrino
  oscillations: the global picture},''{\em Journal of High Energy Physics} {\bf
  2013} (May, 2013)  50}, \href{http://arxiv.org/abs/1303.3011}{{\tt
  arXiv:1303.3011 [hep-ph]}}.

\bibitem{Giovanetti:2024orj}
C.~Giovanetti, M.~Schmaltz, and N.~Weiner, ``{Neutrino-Dark Sector
  Equilibration and Primordial Element Abundances},''
  \href{http://arxiv.org/abs/2402.10264}{{\tt arXiv:2402.10264 [hep-ph]}}.

\bibitem{ACT:2020gnv}
{\bf ACT} Collaboration, S.~Aiola {\em et al.}, ``{The Atacama Cosmology
  Telescope: DR4 Maps and Cosmological Parameters},''
  \href{http://dx.doi.org/10.1088/1475-7516/2020/12/047}{{\em JCAP} {\bf 12}
  (2020)  047}, \href{http://arxiv.org/abs/2007.07288}{{\tt arXiv:2007.07288
  [astro-ph.CO]}}.

\bibitem{SPT-3G:2021vps}
{\bf SPT-3G} Collaboration, J.~A. Sobrin {\em et al.}, ``{The Design and
  Integrated Performance of SPT-3G},''
  \href{http://dx.doi.org/10.3847/1538-4365/ac374f}{{\em Astrophys. J. Supp.}
  {\bf 258} (2022) no.~2, 42}, \href{http://arxiv.org/abs/2106.11202}{{\tt
  arXiv:2106.11202 [astro-ph.IM]}}.

\bibitem{DESI:2016fyo}
{\bf DESI} Collaboration, A.~Aghamousa {\em et al.}, ``{The DESI Experiment
  Part I: Science,Targeting, and Survey Design},''
  \href{http://arxiv.org/abs/1611.00036}{{\tt arXiv:1611.00036 [astro-ph.IM]}}.

\bibitem{DESI:2024lzq}
{\bf DESI} Collaboration, A.~G. Adame {\em et al.}, ``{DESI 2024 IV: Baryon
  Acoustic Oscillations from the Lyman Alpha Forest},''
  \href{http://arxiv.org/abs/2404.03001}{{\tt arXiv:2404.03001 [astro-ph.CO]}}.

\bibitem{DESI:2024mwx}
{\bf DESI} Collaboration, A.~G. Adame {\em et al.}, ``{DESI 2024 VI:
  Cosmological Constraints from the Measurements of Baryon Acoustic
  Oscillations},'' \href{http://arxiv.org/abs/2404.03002}{{\tt arXiv:2404.03002
  [astro-ph.CO]}}.

\bibitem{DESI:2024uvr}
{\bf DESI} Collaboration, A.~G. Adame {\em et al.}, ``{DESI 2024 III: Baryon
  Acoustic Oscillations from Galaxies and Quasars},''
  \href{http://arxiv.org/abs/2404.03000}{{\tt arXiv:2404.03000 [astro-ph.CO]}}.

\end{thebibliography}\endgroup

\newpage
\appendix

\section{Derivation of evolution equations}
\subsection{Background equation} \label{app:background_detailed_balance}
We start from \cref{eq: rho_nu background,eq: rho_ndark background}. The question is now the derivation of the source term. For this we can turn to the Dodelson-Widrow-like addition to the Boltzmann equation of \cref{eq:fundamental}, namely
\begin{equation}
    \frac{\partial f_\nuSM }{\partial \ln a} - p \frac{\partial f_\nuSM}{\partial p} = - \frac{\langle \Gammaph \rangle}{H} (f_\nu - f_\nudark)~.
\end{equation}
It is trivial to recognize that to get an equation for the total energy density one needs to integrate over $g_\nuSM/(2\pi)^3 \cdot \mathrm{d}^3p E$. We remind that we assume massless neutrinos, allowing us to integrate instead over $E \to p$. We immediately recognize that this will give 
\begin{equation}\label{app:eq:background_density}
    \rho_\nuSM = \frac{g_\nuSM}{(2\pi)^3} \int \mathrm{d}^3p \cdot p f_\nuSM = 2\pi^2 g_\nuSM a^{-4}\int \mathrm{d}q\,q^3 f_\nuSM ~,
\end{equation}
where for the last equation we have introduced $q = a p$, which is useful for the remainder of the appendix. We then find from \cref{eq:fundamental} (using partial integration in $p$) that
\begin{equation}
    \frac{\partial \rho_\nuSM}{\partial \ln a} + 4 \rho_\nuSM = - \frac{g_\nu}{(2\pi)^3}  \int \mathrm{d}^3p \cdot p \cdot \frac{\langle \Gammaph \rangle}{H} (f_\nu - f_\nudark) ~.
\end{equation}
Comparing to \cref{eq: rho_nu background} we recognize that
\begin{equation}
    \frac{\partial S}{\partial \ln a} = - \frac{g_\nu}{(2\pi)^3}  \int \mathrm{d}^3p \cdot p \cdot \frac{\langle \Gammaph \rangle}{H} (f_\nu - f_\nudark) ~.
\end{equation}

While solving the above integral for the energy dependent conversion rate $\Gamma(E)$ is highly non-trivial, we use the thermally averaged rate which allows to pull the factor of $\langle \Gammaph \rangle/H$ out of the integral. This simplification unavoidably introduces an important error which we now make explicit. After pulling out the $\langle \Gammaph \rangle/H$, the integration is straight forward and results in
\begin{equation}\label{eq:wrong background}
    \frac{\partial \rho_\nuSM}{\partial \ln a} + 4 \rho_\nuSM \stackrel{?}{=} \frac{\langle \Gammaph \rangle}{H} \left[ \frac{g_\nuSM}{g_\nudark}3R_{3,\nudark}  -  \rho_\nuSM\right]~.
\end{equation}
As long as the equilibration temperature is well above the dark fermion mass $\mdark$, we have $3R_{3,\nudark} \simeq \rho_\nudark$ and $\frac{g_\nuSM}{g_\nudark} \simeq \frac{\rho_\nuSM}{\rho_\nudark}\big|_{\rm eq.}$, and hence the above equation gives the correct result. On the contrary, once thermalization occurs at $\Tequil \simeq \mdark$, detailed balance is violated. We overcome this problem by imposing $\frac{g_\nuSM}{g_\nudark} \to \frac{\rho_\nuSM}{R_{{3,\nudark},{[T \to T_\nu]}}}$, which gives \cref{eq: derivative_density_detailed_balance}, namely
\begin{equation}
    \frac{\partial \rho_\nuSM}{\partial \ln a} + 4\rho_\nuSM  = -\frac{\langle \Gammaph \rangle}{H} \rho_\nuSM\left[1 - \frac{R_{3,\nudark}}{R_{{3,\nudark},{[T \to T_\nu]}}}\right] ~.
\end{equation}

%%%%%%%%%%%%%%

\subsection{SM Neutrinos: Effective addition to the Liouville operator}\label{app:liouville}
In this section we derive the effective addition to the Liuoville operator caused by the temperature change of the active neutrinos. Note that another alternative (simplified) view is presented in \cref{app:simplified_liouville}. The Boltzmann equation for a general particle species can be written as
\begin{equation}
    L[f] \equiv \frac{\mathrm{D}f}{\mathrm{D}\tau} = C[f]~,
\end{equation}
where the left hand side is the so-called Liouville operator, while the right hand side is the collision operator. In this appendix we show that the \enquote*{usual} treatment of constructing the momentum-integrated equations of motion creates the illusion of an additional term in the Boltzmann equation. The left hand side of the Boltzmann equation is given by:
\begin{align}
    L[f] = \frac{Df_\nuSM}{D\tau} =\frac{\partial f_\nuSM}{\partial \tau} + \frac{\mathrm{d}x^i}{\mathrm{d}\tau}\frac{\partial f_\nuSM}{\partial x^i} + \frac{\mathrm{d}q}{\mathrm{d}\tau}\frac{\partial f_\nuSM}{\partial q}  + \frac{\mathrm{d}n^i}{\mathrm{d}\tau}\frac{\partial f_\nuSM}{\partial n^i} ~.
\end{align}
We expand the distribution function into its zeroth order part $f^0_\nuSM$ and a perturbed piece $\Psi_\nuSM$
\begin{align}
    f_\nuSM = f^0_\nuSM\left[1 + \Psi_\nuSM\right]~,
\end{align}
We use also the usual total derivatives to first order (reminding that we are assuming massless neutrinos)
\begin{align}\label{app:eq:total_derivs}
    \frac{\mathrm{d}x^i}{\mathrm{d}\tau} \simeq \hat{n}^i  \quad,\quad \frac{\mathrm{d}n^i}{\mathrm{d}\tau} \simeq 0
    \quad,\quad
    \frac{\mathrm{d}q}{\mathrm{d}\tau} = -q\left[
    - \dot{\phi}_{CN} + n_i\partial_i\psi_{CN}	\right]~.
\end{align}
with $\psi_{CN}$ and ${\phi}_{CN}$ the metric perturbations in conformal Newtonian gauge. We find the usual form of the Liouville operator at first order, namely $L_0[f] = \partial f^0_\nuSM/\partial \tau$. We find the first-order term (also using the Fourier-transform)
\begin{align}\label{app:eq:liouville}
    &L_1[f] = \frac{\partial \left( f^0_\nuSM\Psi_\nuSM \right)}{\partial \tau} + i  \left(\vec{k} \cdot\hat{n} \right) \left( f^0_\nuSM \Psi_\nuSM \right) 
    + q\left[ \dot{\phi}_{CN} -  i \left(\vec{k} \cdot\hat{n} \right) \psi_{CN}	\right] \frac{\partial \left( f^0_\nuSM\right)}{\partial q}~.
\end{align}
In contrast to the typical Boltzmann hierarchy formalism, we will not divide the expression through by the zeroth order distribution function yet. Instead, we first define
\begin{align}
    \widetilde{F}_\nu = \int q^3 dq f^0_\nu \Psi_\nu~,
\end{align}
which can be inserted in the expression of \cref{app:eq:liouville} by integrating over $q^3 \mathrm{d}q$ without problems. Indeed the integral and time-derivatives can be exchanged, since terms like $\mathrm{d} q/\mathrm{d} \tau$ would generate only second-order corrections according to \cref{app:eq:total_derivs}. Note that we also use \cref{app:eq:background_density}. We find
\begin{align}
	& \frac{\partial \widetilde{F}_\nuSM}{\partial \tau} + i k\mu\, \widetilde{F}_\nuSM
	- 4\frac{2\pi^2\rho_\nuSM a^4}{g_\nuSM}\left[ \dot{\phi}_{CN} -  i k\mu\, \psi_{CN}	\right] ~.
\end{align}
where $\mu\equiv\hat{k}\cdot\hat{n}$ as usual. Our last step is to divide everything by $\int dq\,q^3 f^0_\nuSM$ in order to move from $\tilde{F}$ to the usual $F$. The only complication comes from the $\partial_\tau$ term,  which we can be rewritten using
\begin{align}
    \frac{1}{\int dq\,q^3 f^0_\nuSM}\frac{\partial \widetilde{F}_\nuSM}{\partial \tau} = \frac{\partial }{\partial \tau} \frac{\widetilde{F}_\nuSM}{\int dq\,q^3 f^0_\nuSM} - \widetilde{F}_\nuSM\frac{\partial }{\partial \tau} \frac{1}{\int dq\,q^3 f^0_\nuSM} 
	=  \frac{\partial F_\nuSM}{\partial \tau} + F_\nuSM \frac{\partial_\tau (a^4 \rho_\nuSM)}{a^4 \rho_\nuSM}~.
\end{align}
Plugging that in we get
\begin{align}
	& \frac{\partial F_\nuSM}{\partial \tau} + i k\mu\, F_\nuSM 
	- 4\left[ \dot{\phi}_{CN} -  i k\mu\, \psi_{CN}	\right]
     + F_\nuSM \left[\frac{\partial_\tau (\rho_\nuSM a^4)}{\rho_\nuSM a^4} 
	\right]~.
\end{align}
This gives the usual equations in the limit that $\rho_\nu \propto a^{-4}$, and otherwise includes additional corrections due to changes in the comoving temperature
\begin{equation}
	\frac{\mathrm{d}(\rho_\nuSM a^4)}{\mathrm{d}\tau} = 4 (\rho_\nuSM a^4) \frac{\mathrm{d} \ln(aT_\nuSM)}{\mathrm{d}\tau} = a^4\frac{\mathrm{d}S}{\mathrm{d}\tau}~,
\end{equation}
where in the last step we bring back the source term from \cref{eq: rho_nu background}, emphasizing that this additional term derived from the 'usual' treatment of the Liouville equation is equivalent to the background source term (which itself arises from the zeroth-order collision term). In this sense, one can consider this term to be a part of the collision operator and not the Liouville operator. Indeed, if we had not performed the usual transformation of $f_\nuSM \to F_\nu$ we would not have had this term arise except through the collision operator. We therefore can define the 'usual' Liouville part to be
\begin{equation}
	\frac{\partial F_\nu}{\partial \tau} + i k\mu\, F_\nu
	- 4\left[ \dot{\phi}_{CN} -  i k\mu\, \psi_{CN}	\right] ~.
\end{equation}
At this stage, we can construct the Boltzmann hierarchy via the usual multipole expansion 
\begin{equation}
    F_\nu(\vec{k},\hat{n},\tau) = \sum^{\infty}_{\ell=0}(-i)^\ell(2\ell+1) F_{\nu}^\ell(\vec{k},\tau)P_\ell(\hat{k}\cdot\hat{n})~.
    \end{equation}
We can then turn to constructing the collision term for each multipole $F_{\nu}^\ell$ as usual.

%%%%%%%%%%%%%%%%%%%%%%%%%%

\subsection{Another view on background energy injection}\label{app:simplified_liouville}
    In this appendix we give a few more intuitive arguments which give the same results as in \cref{app:liouville}. They are meant to give a slightly deeper understanding of the relation between the Liouville and Collision operator views on the additional terms.

    First, we note that there is additional injection of energy density at the background level, such that neither the active neutrinos nor the dark sector are separately obeying simple conservation equations.  Instead, we have $\mathrm{d}\rho/\mathrm{d}t = -3 H (\rho+P) + \mathrm{d}S/\mathrm{d}t$, so we have to adjust the usual equations whenever a term of the kind $\mathrm{d}\rho/\mathrm{d}t$ appears, using the additional $\mathrm{d}S/\mathrm{d}t$ that would otherwise be neglected. For example, there is the Boltzmann equation for the overdensity, which is:
    \begin{equation}
       \mathrm{d}(\delta \rho)/\mathrm{d}t = \ldots  
    \end{equation}
    Now, the usual Liouville equation for the overdensity can easily be obtained from
    \begin{equation}
       \mathrm{d}(\delta \rho)/\mathrm{d}t = \mathrm{d}\delta/\mathrm{d}t \cdot \rho + \delta \cdot \mathrm{d}\rho/\mathrm{d}t = \rho \cdot \left[\mathrm{d}\delta/\mathrm{d}t - 3 H \delta (1+w) \right] ~,
    \end{equation}
    but instead with the additional term we get 
    \begin{equation}
       \mathrm{d}(\delta \rho)/\mathrm{d}t = \mathrm{d}\delta/\mathrm{d}t \cdot \rho + \delta \cdot \mathrm{d}\rho/\mathrm{d}t = \rho \cdot \left[\mathrm{d}\delta/\mathrm{d}t - 3 H \delta (1+w) + \delta \cdot \mathrm{d}S/\mathrm{d}t /\rho \right] ~.
    \end{equation}
    So for the equation of motion of the overdenisty $\delta$ there will be an additional term of the type $-\delta \mathrm{d}S/\mathrm{d}t/\rho$ on the right-hand side. Indeed, since we have $\mathrm{d}S/\mathrm{d}t =  4 \rho_\nu [\mathrm{d}\ln (aT_\nu)]/\mathrm{d}t$ (from \cref{eq: rho_nu background} and using $\rho_\nu = [\pi^2 g_\nu/30]\cdot T_\nu^4$) we can immediately conclude the additional term for the equation of the overdensity to be
    \begin{equation}
        -\delta_\nu \cdot 4 \rho_\nu \cdot \frac{\mathrm{d}\ln (aT_\nu)}{dt}  \frac{1}{\rho_\nu} = -4 \delta_\nu \cdot \frac{\mathrm{d}\ln (aT_\nu)}{dt}~.
    \end{equation}
    This is indeed the same term as we got before. The same trick we can play for the other moments to get
    \begin{equation}
        \mathrm{d}[(\rho+P)\theta]/dt = \mathrm{d}(1+w)/dt \rho \theta + (\rho+P)\mathrm{d}\theta/\mathrm{d}t + (1+w)\mathrm{d}\rho/\mathrm{d}t \theta =  (\rho+P)\left[\mathrm{d}\theta/\mathrm{d}t + \frac{\mathrm{d}w/\mathrm{d}t}{1+w} \theta + \frac{\mathrm{d}\ln \rho}{\mathrm{d}t} \theta\right]~,
    \end{equation}
    where the additional term is now $-\mathrm{d}S/\mathrm{d}t  \cdot \theta/\rho$, which can be generalized to $-\mathrm{d}S/\mathrm{d}t \cdot  F_\ell/\rho$ , always giving exactly the correct term from our equations above. However, we can also use the trick for the dark sector, for which we will get the additional term $+\mathrm{d}S/\mathrm{d}t \cdot \delta_\mathrm\DS/\rho_\mathrm\DS$ or $+\mathrm{d}S/\mathrm{d}t \cdot  \theta_\mathrm\DS/\rho_\mathrm\DS$. Then we get the additional terms with a prefactor of
    \begin{equation}
        +\mathrm{d}S/\mathrm{d}t /\rho_\mathrm\DS = +\left[4 \rho_\nu \cdot \frac{d\ln (aT_\nu)}{dt}\right] / \rho_\mathrm\DS = + 4 \frac{\rho_\nu}{\rho_\mathrm\DS} \cdot \frac{d\ln (aT_\nu)}{dt}~.
    \end{equation}
    Note that we are getting in total the opposite sign of the neutrino term, such that their sum would cancel out.

 \subsection{The monopole as a temperature fluctuation}
	\label{app: monopole vs temperature}
	In this appendix we derive the relation between an energy density perturbation and a temperature perturbation. Assume that locally our distribution function has a perturbation which is independent of the direction, namely we have local thermal equilibrium but the local temperature somewhat deviates from the averaged temperature. In equations we mean that $T(x) = T + \delta T$ where $T$ is the averaged temperature, and $\delta T$ depends only on $x$. For later convenience let us define $\delta_T \equiv \delta T / T$. We can now find the relation of the temperature perturbation to the energy density perturbation by
	\begin{align}
		\rho[1+\delta] = \rho+\delta \rho = \rho + \delta T \partial_T \rho
		= \rho + \delta_T \frac{\partial \rho}{\partial \ln T} = \rho \left[1 + \frac{3(\rho + R_0)}{\rho}\delta_T\right] ~,
	\end{align}
    where we have used \cite[Eq.~(A.8)]{Schoneberg:2022grr}. By comparing the left and right and side we find the desired formula 
	\begin{align}\label{eq: monopole vs. temperature}
		\delta_T & = \frac{1}{3(1 + R_0/\rho)} \delta 
        \equiv \frac{1}{3(1 + w_{R_0})} \delta
    ~,
	\end{align}
    where we have used the definition of the pseudo equation of state
    \begin{align}\label{app:eq:pseudo_eq_of_state}
        w_{R_0,\xi} \equiv R_{0,\xi} / \rho_\xi~.
    \end{align}
Note that since $\delta_{T_\DS} \stackrel{!}{=} \delta_{T_\nudark}$ we have
    \begin{align}
        \delta_{\nudark} = \frac{1 + w_{R_0,\nudark}}{1 + w_{R_0,DS}} \delta_\DS ~,
    \end{align}
    and that the equilibrium between the dark sector and the active neutrinos would be established for
    \begin{align}
    \delta_{T_\nu} \to \delta_{T_\DS} \quad \Rightarrow \quad
        \delta_\nuSM \to \frac{1 + w_{R_0,\nuSM}}{1 + w_{R_0,DS}} \delta_\DS ~.
    \end{align}
Note that in principle the $\delta_\DS$ does not relate to a single species but rather the sum of two. However, this does not pose a problem, as we demonstrate below. First, we remind that
\begin{equation}
    \delta_\mathrm\DS = \frac{\delta \rho_\mathrm\DS}{\rho_\mathrm\DS} = \frac{\delta \rho_\phi + \delta \rho_\nudark}{\rho_\phi+\rho_\nudark} = \frac{\rho_\phi \delta_\phi + \rho_\nudark \delta_\nudark}{\rho_\phi+\rho_\nudark} ~.
\end{equation}
Furthermore, we can relate $\nudark$ to $\phi$ (both being single species) using \cref{eq: monopole vs. temperature}, giving
\begin{align}
        \delta_{\nudark} = \frac{1 + w_{R_0,\nudark}}{1 + w_{R_0,\phi}} \delta_{\phi} ~,
    \end{align}
which we can plug in to get 
\begin{equation}
    \delta_\mathrm\DS  = \frac{\rho_\phi \frac{1+w_{R_0,\phi}}{1+w_{R_0,\nudark}} \delta_\nudark + \rho_\nudark \delta_\nudark}{\rho_\phi+\rho_\nudark} = \frac{\rho_\phi \frac{1+w_{R_0,\phi}}{1+w_{R_0,\nudark}} + \rho_\nudark}{\rho_\phi+\rho_\nudark} \cdot \delta_\nudark ~.
\end{equation}
By using $R_{0,\mathrm\DS} = R_{0,\phi} + R_{0,\nudark}$ (by linearity of the integrals), we get
\begin{equation}
    w_{R_0,\mathrm\DS} = \frac{R_{0,\phi}}{\rho_\phi+\rho_\nudark} + \frac{R_{0,\nudark}}{\rho_\phi+\rho_\nudark} = w_{R_0,\phi} b + (1-b) w_{R_0,\nudark} ~,
\end{equation}
where $b = \frac{\rho_\phi}{\rho_\phi+\rho_\nudark}$\,. Then we can reformulate the above to get 
\begin{equation}
    \delta_\mathrm\DS  = \frac{\rho_\phi \frac{1+w_{R_0,\phi}}{1+w_{R_0,\nudark}} + \rho_\nudark}{\rho_\phi+\rho_\nudark} \cdot \delta_\nudark = \left[b \frac{1+w_{R_0,\phi}}{1+w_{R_0,\nudark}} + (1-b) \right] \delta_\nudark ~,
\end{equation}
which can be simplified to give
\begin{equation}
    \delta_\mathrm\DS  =  \left[b +b w_{R_0,\phi} + (1-b) 
 + (1-b) w_{R_0,\nudark}\right]\frac{\delta_\nudark}{1+w_{R_0,\nudark}} = \left[1+w_{R_0,\mathrm\DS}\right]\frac{\delta_\nudark}{1+w_{R_0,\nudark}} ~,
\end{equation}
so the result was well justified.
	
	%%%%%%%%%%%%%%%%%%%
	%%%%%%%%%%%%%%%%%%%
	%%%%%%%%%%%%%%%%%%%

%%%%%%%%%%%%
\newpage
\section{Triangle plots and Tables}\label{app:tables}

In this section we provide additional triangle plots and tables that summarize our results.

\begin{table}[H]
    \centering
    \begin{tabular}{c c|c c c c c c}
        Model & Dataset & $H_0$ & $\Omega_m h^2$& $n_s$ & $\log_{10}(\theta_0)$ & $\log_{10}(\alphad)$ &  $\log_{10}(\mdark/ \rm eV)$\\ \hline \rule{0pt}{1em}
        \multirow{2}{*}{$\Lambda$CDM} & baseline & $67.84 \pm 0.41$ & $0.14232 \pm 0.00088$ & $0.9641 \pm 0.0037$ & --- & --- &--- \\
        & +SH0ES & $68.81 \pm 0.39$ & $0.14059 \pm 0.00080$ & $0.9691 \pm 0.0037$ & --- & --- &---\\ \hline \rule{0pt}{1em}
        \multirow{2}{*}{S\wzdrnu (narrow prior)}  & baseline & $67.86 \pm 0.41$ &${0.14246}^{+0.00089}_{-0.00092}$ & ${0.9651}^{+0.0040}_{-0.0043}$ & $< -13$ & ---& Unconstrained \\
        & +SH0ES & $68.89 \pm 0.41$ & $0.14096 \pm 0.00098$ & ${0.9722}^{+0.0061}_{-0.0056}$ &  $< -13$ & --- & Unconstrained\\ \hline \rule{0pt}{1em}
        S\wzdrnu ($1\nu$ interaction) & baseline & $67.90 \pm 0.42$ & $0.14254 \pm 0.00093$ & $0.9654 \pm 0.0043$ & $< -13$ & --- & Unconstrained\\ \hline \rule{0pt}{1em}
        S\wzdrnu (broad prior) & baseline & $67.84 \pm 0.41$ & $0.14232 \pm 0.00085$ & $0.9641 \pm 0.0037$ & $< -11$ & --- & Unconstrained \\
        S\wzdrnu (IDR limit) & baseline & $69.64 \pm 0.44$ & $0.14735 \pm 0.00092$ & $0.9559 \pm 0.0045$ & $-10.4 \pm 1.1$ & --- & ${3.01}^{+0.64}_{-0.54}$ \\
        S\wzdrnu (multimodal) & baseline & ${69.17}^{+0.74}_{-0.89}$ & ${0.1460}^{+0.0019}_{-0.0025}$ & $0.9566 \pm 0.0055$ & $-11.6 \pm 1.7$ & --- & $> 1.5$ \\ \hline \rule{0pt}{1em}
        \multirow{2}{*}{Anomaly} & baseline & $68.15 \pm 0.41$ & $0.14293 \pm 0.00091$ &  $0.9590 \pm 0.0059$ & --- & $-13.12 \pm 0.35$ & ---\\
        & +SH0ES & $69.14 \pm 0.41$ & $0.14146 \pm 0.00088$ & $0.9613 \pm 0.0074$ & --- & ${-13.29}^{+0.47}_{-0.44}$ & ---\\ \hline \rule{0pt}{1em}
        \multirow{2}{*}
        {\wzdrnu (narrow prior)} & baseline & $68.19 \pm 0.49$ &$0.1434 \pm 0.0012$ & $0.9643 \pm 0.0062$ & $< -1.2$ & ${-13.3}^{+1.2}_{-1.0}$ & ${0.11}^{+0.48}_{-0.59}$  \\
        & +SH0ES & $69.63 \pm 0.49$& $0.1440 \pm 0.0015$ & $0.9738 \pm 0.0047$ & $< -1.5$& ${-12.72}^{+0.52}_{-0.44}$ & ${0.60}^{+0.14}_{-0.11}$\\ \hline \rule{0pt}{1em}
        \multirow{2}{*}{\wzdrnu} & baseline & $67.90 \pm 0.42$ & $0.14244 \pm 0.00088$ & $0.9639 \pm 0.0038$ & $< -1.3$ & $< -11$ & Unconstrained \\
        & +SH0ES & $70.53 \pm 0.41$ & ${0.14653}^{+0.00093}_{-0.00095}$ & ${0.9494}^{+0.0050}_{-0.0048}$ &  $< -1.6$ & $-9.84 \pm 0.93$ & $1.77 \pm 0.13$ \\ \hline \rule{0pt}{1em}
        \multirow{2}{*}{\wzdrnu ($1\nu$ interaction)} & baseline & $68.00 \pm 0.46$ &$0.14268 \pm 0.00099$ & $0.9633 \pm 0.0040$ & $< -2.0$ & $< -9.7$ & $> 0.40$ \\
        & +SH0ES &$69.38 \pm 0.42$& ${0.14273}^{+0.00100}_{-0.00095}$ &$0.9619 \pm 0.0051$ &  $< -1.1$ & $-9.8 \pm 1.3$ & $> 0.72$ \\ \hline
    \end{tabular}
    \caption{Mean and $1\sigma$ uncertainty (or $95\%$ upper limit) for selected parameters in all the cases considered in this work. We shorten the \wzdrnu model with $\alphad=1$ to S\wzdrnu.}
    \label{tab:posteriors}
\end{table}

\begin{table}[H]
    \centering
    \begin{tabular}{c c|c c c c c c}
        Model & Dataset & $H_0$ & $\Omega_m h^2$& $n_s$ & $\log_{10}(\theta_0)$ & $\log_{10}(\alphad)$ &  $\log_{10}(\mdark/\rm eV)$\\ \hline \rule{0pt}{1em}
        \multirow{2}{*}{$\Lambda$CDM} & baseline & 67.80 & 0.1424 & 0.965 & --- & --- & ---\\
        & +SH0ES & 68.8 & 0.1405 & 0.969 & --- & --- & ---\\ \hline \rule{0pt}{1em}
        \multirow{2}{*}{S\wzdrnu (narrow prior)}  & baseline & 67.88 & 0.1422 & 0.965 & -14.5 & --- & 0.53\\
        & +SH0ES & 67.94 & 0.1405 & 0.967 & -14.0 & --- & 0.41
        \\ \hline \rule{0pt}{1em}
        S\wzdrnu ($1\nu$ interaction) & baseline & 67.89 & 0.1429 & 0.969 & -13.2 & --- & 0.53
        \\ \hline \rule{0pt}{1em}
        S\wzdrnu (broad prior) & baseline & 67.93 & 0.1421 & 0.965 & -13.4 & --- & -2.97\\
        S\wzdrnu (IDR limit) & baseline & 68.47 & 0.1436 & 0.956 & -13.1 & --- & 2.42\\
        S\wzdrnu (multimodal) & baseline & 69.61 & 0.1477 & 0.954 & -10.3& --- & 2.50
        \\ \hline \rule{0pt}{1em}
        \multirow{2}{*}{Anomaly} & baseline & 68.13 & 0.1429 & 0.961 & --- & -13.18 & --- \\
        & +SH0ES & 69.06 & 0.1418 & 0.957 & --- & -13.58 & --- \\ \hline \rule{0pt}{1em}
        \multirow{2}{*}{\wzdrnu (narrow prior)} & baseline & 68.44 & 0.1434 & 0.968 & -1.74 & -12.56 & 0.435\\
        & +SH0ES &  69.77 & 0.1445 & 0.976 & -1.96 & -13.12 & 0.635\\ \hline \rule{0pt}{1em}
        \multirow{2}{*}
        {\wzdrnu} & baseline & 67.88 & 0.1428 & 0.965 & -3.46 & -15.25& 1.695\\
        & +SH0ES &  70.61 & 0.1469 & 0.953 & -1.96 & -10.43 & 1.895 \\ \hline \rule{0pt}{1em}
        \multirow{2}{*}{\wzdrnu ($1\nu$ interaction)} & baseline & 67.83 & 0.1425 & 0.966 & -3.57 & -10.73 & 1.895\\
        & +SH0ES & 69.39 & 0.1438 & 0.962 & -3.03 & -09.30& 0.805\\ \hline
    \end{tabular}
    \caption{Same as \cref{tab:posteriors}, but listing instead the bestfit parameters for each of the cases. The corresponding likelihoods are reported in \cref{tab:chisquares}.}
    \label{tab:bestfits}
\end{table}

 \begin{figure}[H]
     \centering
     \includegraphics[width=0.7\textwidth]{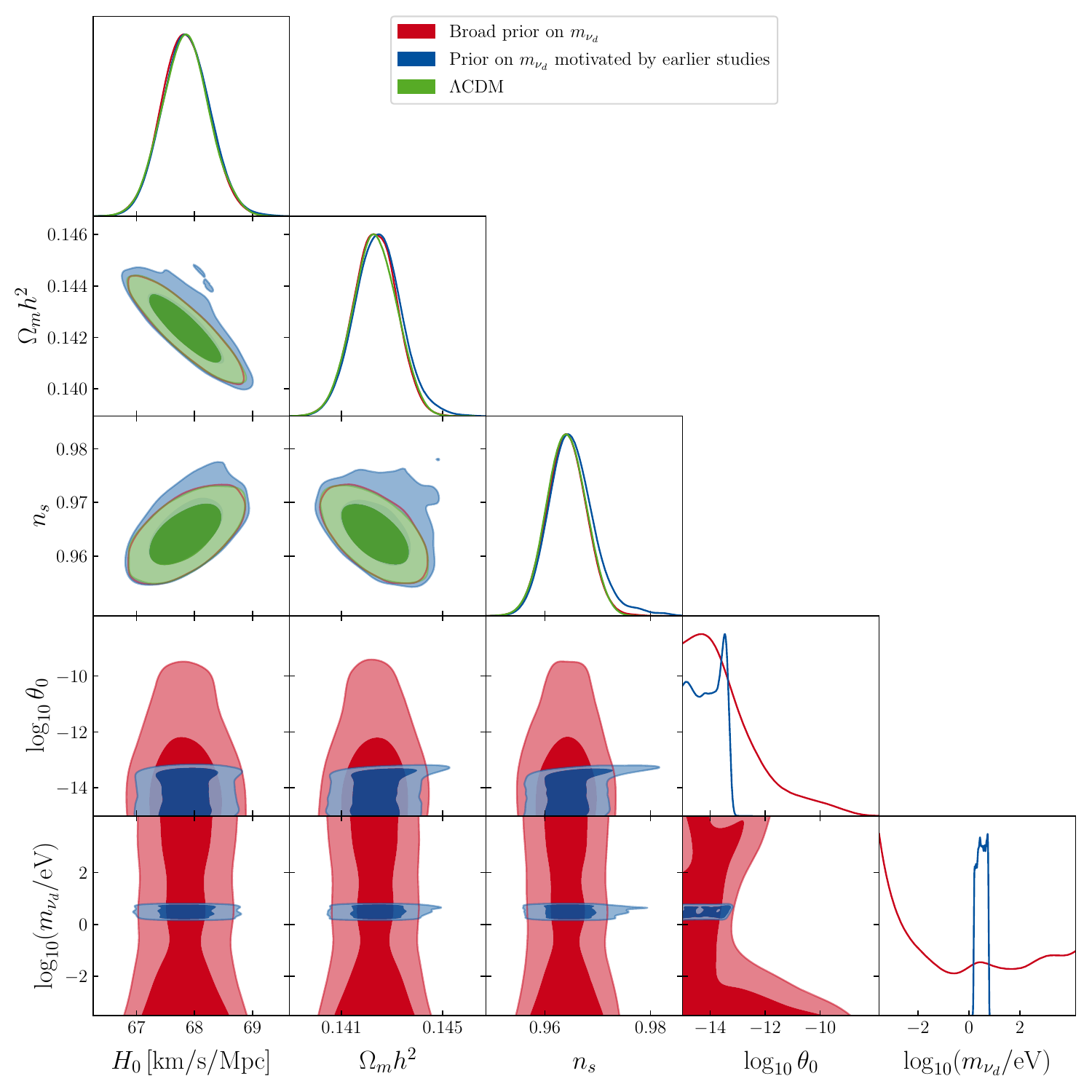}
     \caption{\label{fig:extended_range} One-dimensional posteriors and two-dimensional constraints at 68\% and 95\% CL for a range of parameters. Shown are the $\Lambda$CDM model and the \wzdrnu model (with $\alphad=1$) with a narrow and wide parameter range for the dark fermion mass ($\mdark$, equivalent to $z_t$) using the baseline dataset.}
 \end{figure}
    
\end{document}